\documentclass{pasj00}
\usepackage{color} 

\begin{document}
\SetRunningHead{Sakuma et al.}{Suzaku observations of the Centaurus cluster}
\Received{2011/06/17}
\Accepted{2011/08/09}

\title{Suzaku Observations of Metal Distributions in the Intracluster 
Medium of the Centaurus Cluster}

\author{Eri \textsc{sakuma}\altaffilmark{1}, 
Naomi \textsc{ota}\altaffilmark{2}, 
Kosuke \textsc{sato}\altaffilmark{1}, 
Takuya \textsc{sato}\altaffilmark{1}, 
and Kyoko \textsc{matsushita}\altaffilmark{1}}
\altaffiltext{1}{Department of Physics, Tokyo University of Science, 
1-3 Kagurazaka, Shinjyuku-ku, Tokyo 162-8601}
\email{j1210617@ed.kagu.tus.ac.jp}
\altaffiltext{2}{Department of Physics, Nara Women's University,
Kitauoyahigashi-machi, Nara 630-8506}


\maketitle

\begin{abstract}
We report the first observations of metal distributions 
in the intracluster medium of the Centaurus cluster 
up to $\sim 0.17~r_{180}$ with Suzaku.
Radial profiles of the O, Mg, Si, S, Ar, Ca, and Fe 
were determined at the outer region of the cluster, 
and their variations appear to be similar to each other.  Within the 
cool core region ($r<0.045~r_{180}$), all the metal distributions 
sharply increased toward the center.  In the central region 
($r<0.015~r_{180}$), the abundances of Si, S, Ar, Ca, and Fe were 
1.5--1.8 solar, while those of O and Mg were approximately 1 solar. 
The derived abundance ratios of O and Mg to Fe were slightly 
lower than those of a set of other clusters. In contrast, the 
calculated mass-to-light ratios (MLRs) for O, Mg, and Fe were 
larger than those of the other clusters.  For the outer region 
of the cool core ($r>0.07~r_{180}$), all the abundances were 
almost constant at $0.5$ solar.  The derived MLRs were 
comparable to those of the other clusters.  This suggests 
that the cD galaxy of the Centaurus cluster efficiently supplies more Fe 
than the other clusters.

\end{abstract}

\section{Introduction}

The metal abundances of the intracluster medium (ICM) provide 
important information regarding the chemical history and evolution 
of clusters.  The metals in the ICM were primarily produced by 
supernovae (SNe) in early-type galaxies \citep{Arnaud1992, Renzini1993}. 
Si and Fe are synthesized both in SNe Ia and SNe II, while O and Mg 
are synthesized predominantly in SNe II. O and Mg
abundances, therefore, are crucial for understanding the history of massive 
stars in clusters of galaxies.

Because the Fe--K lines are prominent in the spectra of the ICM,
the Fe abundances of the ICM have been studied in detail.
ASCA first showed Fe distribution in the ICM 
\citep{Fukazawa1998, Fukazawa2000, Finoguenov2000, Finoguenov2001, 
Ezawa1997}.  \citet{DeGrandi2002} derived Fe distribution and 
its gradient from Beppo-SAX observations. Recently XMM-Newton and 
Chandra observations have been used to study spatial distribution of Fe up to 
$0.3$--$0.4~r_{180}$ \citep{Vikhlinin2005, Baldi2007, Maughan2008,
Leccardi2008, Matsushita2011}.  Within the cool core regions, 
Fe abundance in clusters decreases sharply toward the outer region.  
\citet{Matsushita2011} found that outside the cool cores, both cD 
and non-cD clusters have similar Fe abundance profiles, which are 
flatter than expected from numerical simulations. A simple 
explanation is that a significant fraction of Fe was synthesized
in an early phase of cluster evolution.

XMM-Newton provided the means to study the O and Mg abundances
in the brightest cool cores of clusters and groups of galaxies
\citep{Finoguenov2002, Matsushita2003, Tamura2003, Matsushita2007b}.
However, a higher background level and a strong instrumental Al line 
of the MOS detector onboard XMM-Newton can cause problems in measuring 
O and Mg abundances outside cool cores. The X-ray Imaging 
Spectrometer (XIS) instrument \citep{Koyama2007} onboard Suzaku 
\citep{Mitsuda2007} 
offered an improved line spread function below 1 keV coupled with a 
lower background level. With the Suzaku satellite, O and Mg abundances 
of the ICM outside the cool cores of several clusters and groups of 
galaxies were measured up to 0.2--0.3 $r_{180}$
\citep{Matsushita2007a, Komiyama2009, Sato2007a, Sato2008, Sato2009a, 
Sato2009b, Sato2010}.  Within the cool cores, the derived abundance 
patterns of O/Mg/Si/Fe are comparable to the solar ratio, obtained by adopting 
the solar abundance by \citet{Lodders2003}.  Combining the Suzaku 
results with SNe nucleosynthesis models, \citet{Sato2007b} showed 
the number ratios of SNe II to Ia to be $\sim$3.5, and that Fe was 
synthesized mainly by SNe Ia.  A slight increase in O/Fe and Mg/Fe 
with increasing radius was detected in Abell~1060 \citep{Sato2007a} and 
AWM~7 \citep{Sato2008}, although the error bars were reasonably large. 
MLR, the ratios of metal mass 
in the ICM to the total galaxy (stellar) luminosity, is important for 
studying the origin of metals, because metals are synthesized in 
stars. The integrated values of Fe mass-to-light ratio (IMLR) 
profiles of the $kT=$2--3 keV clusters, such as Abell~262, and AWM~7 
observed with Suzaku increased outward from 0.1~$r_{180}$ to 
0.3--0.4~$r_{180}$.  These results imply that Fe in the ICM 
extends farther than stars in the outer regions. These 
results also indicate early metal enrichment process in the ICM.

The Centaurus cluster (Abell~3526) is a nearby X-ray bright cluster
with a cool core \citep{Allen1994,Ikebe1999,Sanders2008,Takahashi2009} 
at a redshift of $z=0.0104$\@.  At the cluster center, the
 cD galaxy NGC~4696 is associated with the low-power radio source 
PKS~1246-410 \citep{Taylor2006}. The cluster has a plume-like 
structure around the cD galaxy.  
The cluster has a subcluster structure Cen~45 at southeast of cluster center.
With ASCA observations, the Cen~45 region has a hotter temperature 
than surrounding regions \citep{Furusho2001}.
The Centaurus cluster has the highest 
central Fe abundance of the ICM of $\sim$2 solar among nearby cool 
core clusters \citep{DeGrandi2009, Matsushita2011}.  From a Chandra 
observation of the Centaurus cluster \citep{Sanders2006}, 
the abundances of O, Ne, Mg, Si, S, Ar, Ca, Fe, and Ni were studied 
up to 0.035~$r_{180}$.  With XMM-Newton, the abundances of O, Si, 
and Fe up to 0.1~$r_{180}$ and abundance of Mg within 0.03~$r_{180}$ 
were derived \citep{Matsushita2007b}.  The central abundances of 
O and Mg are a factor of two lower than those of Si and Fe.  
These results indicate that Fe in the central region is mainly 
supplied by the ejecta of SN Ia in the cD galaxy.

In this paper, we report the abundance pattern of O/Mg/Si/S/Ar/Ca/Fe 
of the Centaurus cluster observed with Suzaku up to 0.17~$r_{180}$ 
within and outside the cool core. We adopt the solar abundance 
values given by \citet{Lodders2003}, in which the solar Fe abundance 
relative to H is 2.95$\times10^{-5}$ in number.  We use the Hubble 
constant $H_{0}=70$ km s$^{-1}$ Mpc$^{-1}$\@.  The luminosity distance 
to the Centaurus cluster is $D_{\rm L}=44.9$ Mpc, and 1$'$ corresponds 
to 13 kpc\@.  The virial radius 
$r_{180}=1.95h^{-1}_{100}\sqrt{\langle kT \rangle / 10~{\rm keV}}$ 
\citep{Markevitch1998, Evrard1996} is approximately 1.74 Mpc
for the average temperature of $\langle kT \rangle=$ 3.88 keV 
with ASCA \citep{Furusho2001}.  Unless otherwise specified, 
the errors are within the 90\% confidence region for a single 
parameter of interest.

\section{Observation and Data Reduction}

\begin{table*}[t]
\begin{center}
\caption{Suzaku observations of the Centaurus cluster and the background
 region.}
\label{tb:obsinfo}
\begin{tabular}{llllll}
\hline
 Target name & Sequence & Date & \multicolumn{2}{l}{Coordinates (J2000.0)} & Exposure \\
 & Number & & $\alpha$ (degree) & $\delta$ (degree) & (ks) \\
\hline
CENTAURUS\_CLUSTER & 800014010 & 2005-Dec-27 & \timeform{192D2054} & \timeform{-41D3111} &  36.3 \\
CENCL\_OFFSET1 & 800015010 & 2005-Dec-28 & \timeform{192D2054} & \timeform{-41D4440} & 44.7 \\
CENCL\_OFFSET2 & 800016010 & 2005-Dec-29 & \timeform{192D2054} & \timeform{-41D1780} & 43.0 \\
CEN45 & 802008010 & 2007-Dec-24 & \timeform{192D5119} & \timeform{-41D3865} & 58.2 \\
\hline\hline
LOCKMANHOLE & 100046010 & 2005-Nov-14 & \timeform{163D4063} & \timeform{57D6108} & 77.0 \\
LOCKMANHOLE & 101002010 & 2006-May-17 & \timeform{162D9366} & \timeform{57D2557} & 80.4 \\
LOCKMANHOLE & 102018010 & 2007-May-03 & \timeform{162D9257} & \timeform{57D2581} & 96.1 \\
LOCKMANHOLE & 103009010 & 2008-May-18 & \timeform{162D9369} & \timeform{57D2546} & 83.4 \\
\hline
\hline
\end{tabular}
\end{center}
\end{table*}

\begin{figure}[t]
\begin{center}
\FigureFile(80mm,80mm){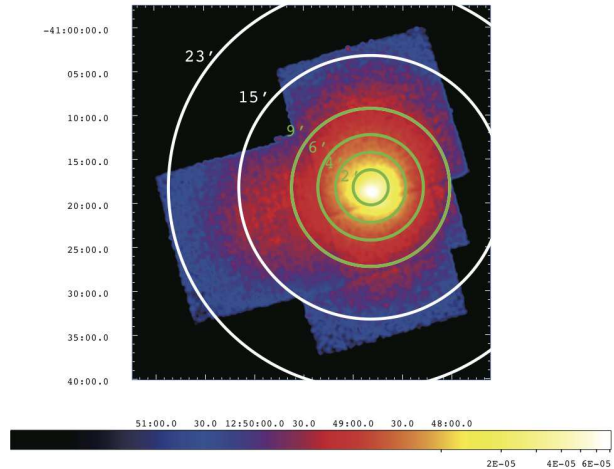}
\caption{XIS~0 image (0.5--4 keV) of the Centaurus cluster.
The images obtained with the four pointing observations  are superposed.
The exposure was corrected, but the NXB and CXB were not subtracted.  
The image was smoothed by a Gaussian of $\sigma=$10 pixels.  
The unit of color bar is counts pixel$^{-1}$ s$^{-1}$\@.  The circles 
indicate the annular regions as mentioned in the spectral analysis.}
\label{fig:xisimage}
\end{center}
\end{figure}

One offset and three central regions of the Centaurus cluster 
were observed in December 2005 and 2007 by Suzaku.  The observed 
regions range to 23$'$, which corresponds to $\sim0.17~r_{180}$, 
from the cD galaxy NGC~4696 at the center.  
The CEN45 field corresponds to the subcluster, the Cen~45 region.
The observation logs are 
given in table~\ref{tb:obsinfo}, and an XIS
 image in 0.5--4.0 keV is shown in figure~\ref{fig:xisimage}.
We used only the XIS data in this study.  The XIS instrument consists 
of four sets of X-ray CCD (XIS~0, 1, 2, and 3).  XIS~1 is a 
back-illuminated (BI) sensor, while XIS~0, 2, and 3 are front-illuminated 
(FI) sensors.  XIS~2, however, has not been available since November 2006;  
hence, the CEN 45 observation was carried out only by XIS~0, 1, 
and 3\@.  The XIS was operated in the normal clocking mode with
 standard 5$\times$5 or 3$\times$3 editing modes.

The analysis was performed with HEAsoft version 6.7 and XSPEC 12.5.1\@.
We used the standard data selection 
\footnote{http://www.astro.isas.ac.jp/suzaku/process/v2changes/
criteria\_xis.html}, and the CALDB files version 2009-04-02\@.  
Because the energy resolution slowly degraded after the launch 
because of radiation damage, this effect was included in the 
redistribution matrix file generated by the ``xisrmfgen'' Ftools task.  
We generated an Ancillary Response File (ARF) for the spectrum of 
each annular sky region, which assumed $r=25'$ circle size of 
the surface brightness profile, which is a sum of two $\beta$-models,
$\beta_{1}=$0.57 ,$r_c=$\timeform{7.3'} and $\beta_{2}=$0.92, 
$r_{c,2}=$\timeform{0.92'} derived from ASCA and ROSAT results 
\citep{Ikebe1999} by the ``xissimarfgen'' Ftools task 
\citep{Ishisaki2007}.
We also included the contamination effect on the optical blocking filter
 of the XIS 
in the ARFs.  We employed the night Earth database 
generated by the ``xisnxbgen'' Ftools task to subtract non-X--ray 
background (NXB).

\begin{table*}[t]
\caption{Results of spectral fits of the Lockman hole region for 
the estimation of the CXB component and the Centaurus offset regions  
for the Galactic components.}
\label{tb:bgd}
\begin{center}
\begin{tabular}{lllll}
\hline
CXB & & & & \\
Seq. Number & $\Gamma$ & $Norm^\ast$ & $\chi^{2}\slash$d.o.f. & \\
\hline
100046010 & 1.53$_{-0.03}^{+0.03}$ & 14.2$_{-0.5}^{+0.6}$ & 1120$\slash$945 & \\ 
101002010 & 1.41$_{-0.04}^{+0.04}$ & 9.3$_{-0.4}^{+0.4}$  & 831$\slash$765 & \\ 
102018010 & 1.43$_{-0.05}^{+0.05}$ & 8.8$_{-0.4}^{+0.4}$  & 662$\slash$617 & \\ 
103009010 & 1.49$_{-0.05}^{+0.05}$ & 9.0$_{-0.5}^{+0.5}$  & 670$\slash$529 & \\ 
\hline\hline
Galactic & & & & \\
OFF1 \& OFF2 & & & & \\
\hline
$kT_{\rm MWH}$ & $Norm_{\rm MWH}^{\dagger}$ & $kT_{\rm LHB}$ & 
$Norm_{\rm LHB}^{\dagger}$ & $\chi^{2}$/d.o.f \\
(keV) & ($\times10^{-3}$) & (keV) & ($\times10^{-3}$) & \\  
\hline
0.16$_{-0.03}^{+0.03}$ & 3.1$_{-1.3}^{+1.7}$ & 0.1 (fixed) & 
0.0$_{-0.0}^{+1.3}$ & 535/561 \\ 
\hline\\[-1ex]
\multicolumn{5}{l}{\parbox{0.6\textwidth}{\footnotesize 
\footnotemark[$\ast$]
The units of photons cm$^{-2}$ s$^{-1}$ sr$^{-1}$ keV$^{-1}$ 
at 1 keV\@.}}\\
\multicolumn{5}{l}{\parbox{0.6\textwidth}{\footnotesize 
\footnotemark[$\dagger$]
Normalization of the apec component
divided by the solid angle, $\Omega^{\makebox{\tiny\sc u}}$,
assumed in the uniform-sky ARF calculation (20$'$ radius),
$Norm = \int n_{\rm e} n_{\rm H} dV \,/~\,[4\pi\,(1+z)^2 
D_{\rm A}^{~2}] \,/\, \Omega^{\makebox{\tiny\sc u}}$ 
$\times 10^{-14}$ 
cm$^{-5}$~400$\pi$~arcmin$^{-2}$, 
where $D_{\rm A}$ is the angular distance to the source.}}\\
\end{tabular}
\end{center}
\end{table*}

\section{Spectral Analysis}

\subsection{Estimation of the Cosmic X-ray Background and Galactic Foreground}

We estimated the cosmic X-ray background (CXB) using four Lockman 
hole observations with Suzaku as shown in table~\ref{tb:obsinfo}.  
Because the observed emissions in our observation were dominated 
by the ICM emission from the Centaurus cluster, we could not estimate 
the CXB level in our observation.  Thus, we used the Lockman hole 
data for the background estimation.  We fitted the spectra from 
the Lockman hole data with an absorbed power-law model.  
The resultant parameters are shown in table~\ref{tb:bgd}.  
The weighted average of photon index $\Gamma~=1.47$, and Norm$=9.9$ photons 
cm$^{-2}$ s$^{-1}$ sr$^{-1}$ keV$^{-1}$ at 1 keV\@ were used.  

The Galactic emissions mainly arise from the local hot bubble (LHB) 
and the Milky Way halo (MWH).  We used a two-temperature model ($apec$;
\cite{Smith2001}) for these emissions. 
The version 1.3.1 of APEC code is used through this paper.
For the outermost region 
($15'$--$23'$) in the offset 1 and 2 regions, we fitted the spectra 
with the model formula 
$apec_{\rm LHB}+wabs*(apec_{\rm MWH}+power$-$law_{\rm CXB}+apec_{\rm
ICM})$\@ as described in \citet{Sato2007a} and \citet{Komiyama2009}.  Here we 
fixed the LHB temperature at 0.1 keV\@ that was typical value of XMM result \citep{Lumb2002}, while 
the MWH temperature was a free parameter. The normalizations of the 
LHB and MWH were allowed to vary.  We also assumed a solar abundance and 
zero redshift for the LHB and MWH components.  The resultant parameters 
are shown in table~\ref{tb:bgd}.  The derived values 
were consistent with those of \citet{Yoshino2009}.

Using these parameters, we generated mock spectra including 
the CXB, the LHB, and the MWH components as the background.  In the spectral 
analysis of the inner (0$'$--$15'$) regions, we subtracted the mock 
spectra as the background from the observed spectra for each sensor.
The CXB and Galactic components in 15$'$-23$'$ region were included 
as models in the spectral fits.  As a result, the 
temperature and abundances in the ICM did not change within the 
statistical errors when we examined the systematic errors of 
our results by changing the background normalization by $\pm$10\% 
as mentioned in subsection \ref{subsec:uncertainties}. 

\begin{figure*}[t]
\begin{minipage}{0.33\textwidth}
\FigureFile(55mm,55mm){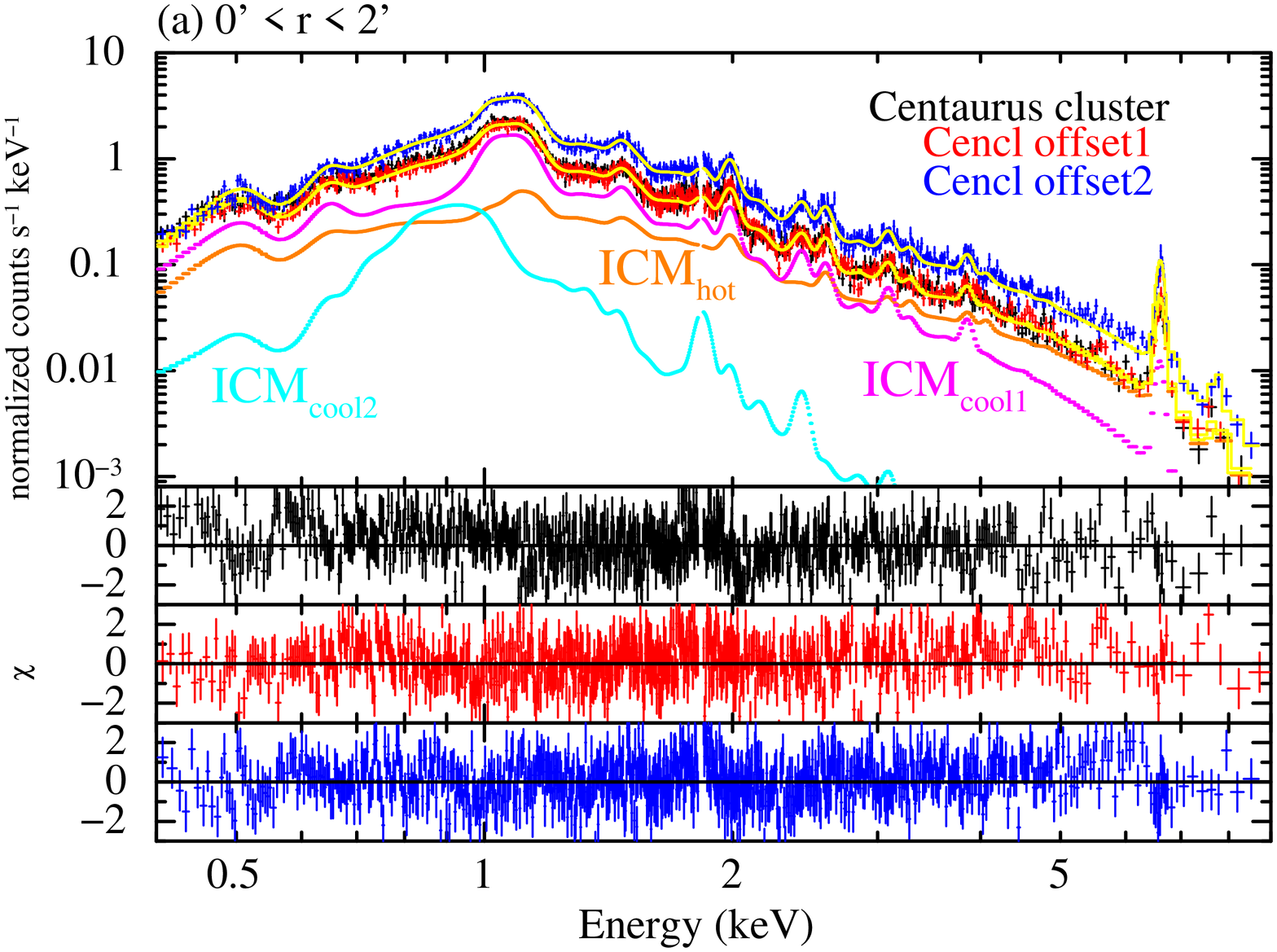}
\end{minipage}\hfill
\begin{minipage}{0.33\textwidth}
\FigureFile(55mm,55mm){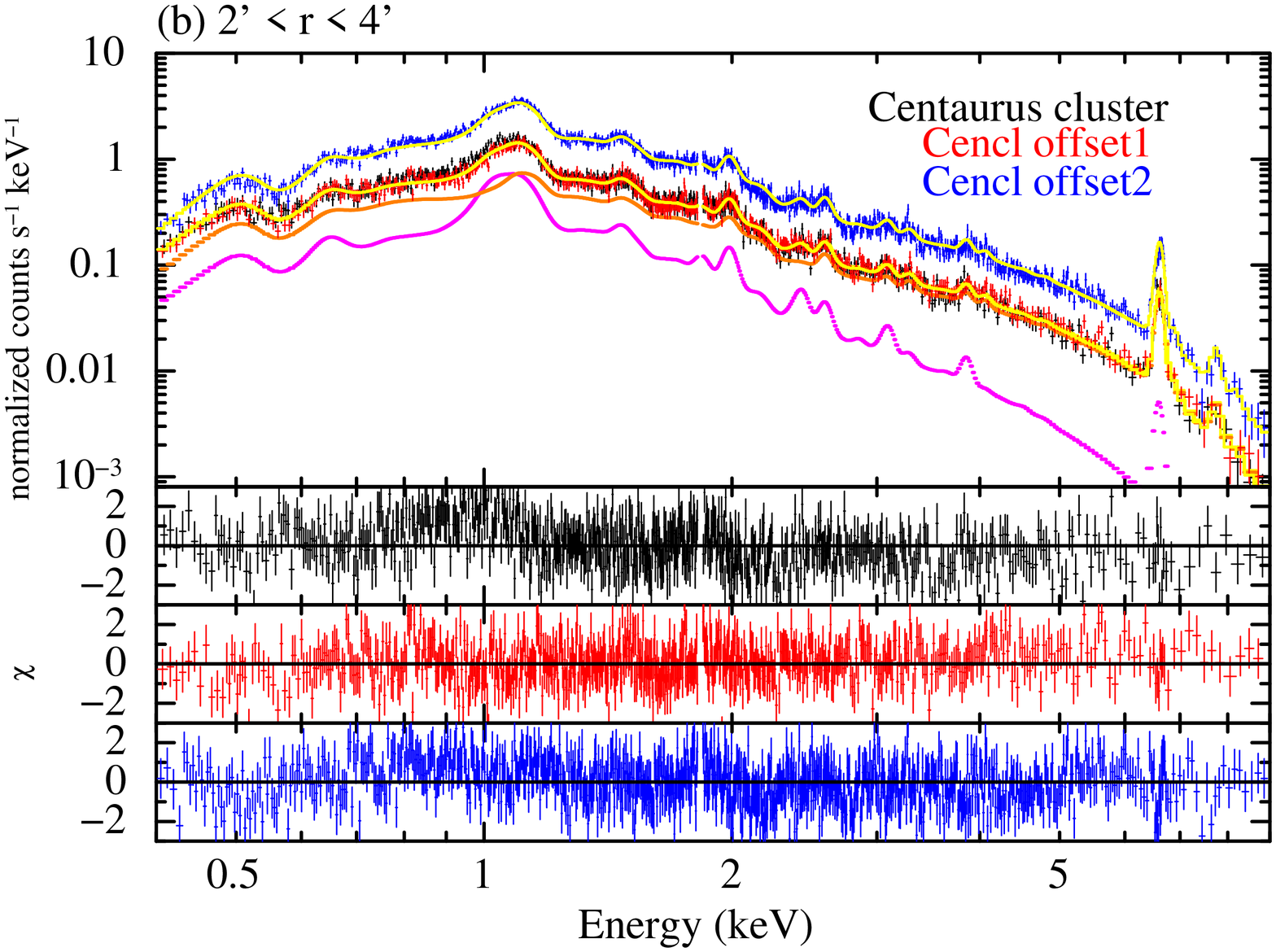}
\end{minipage}\hfill
\begin{minipage}{0.33\textwidth}
\FigureFile(55mm,55mm){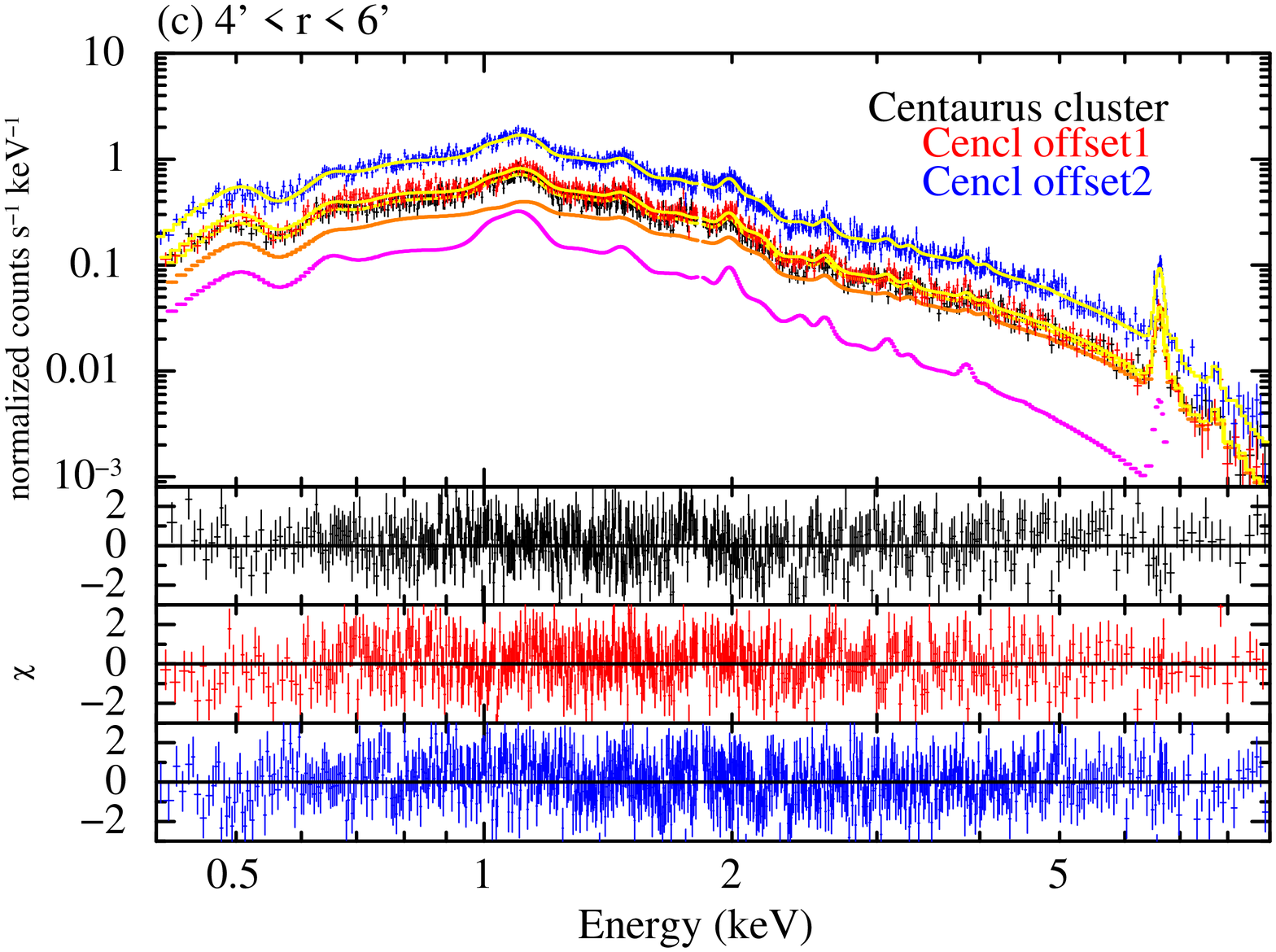}
\end{minipage}

\begin{minipage}{0.33\textwidth}
\FigureFile(55mm,55mm){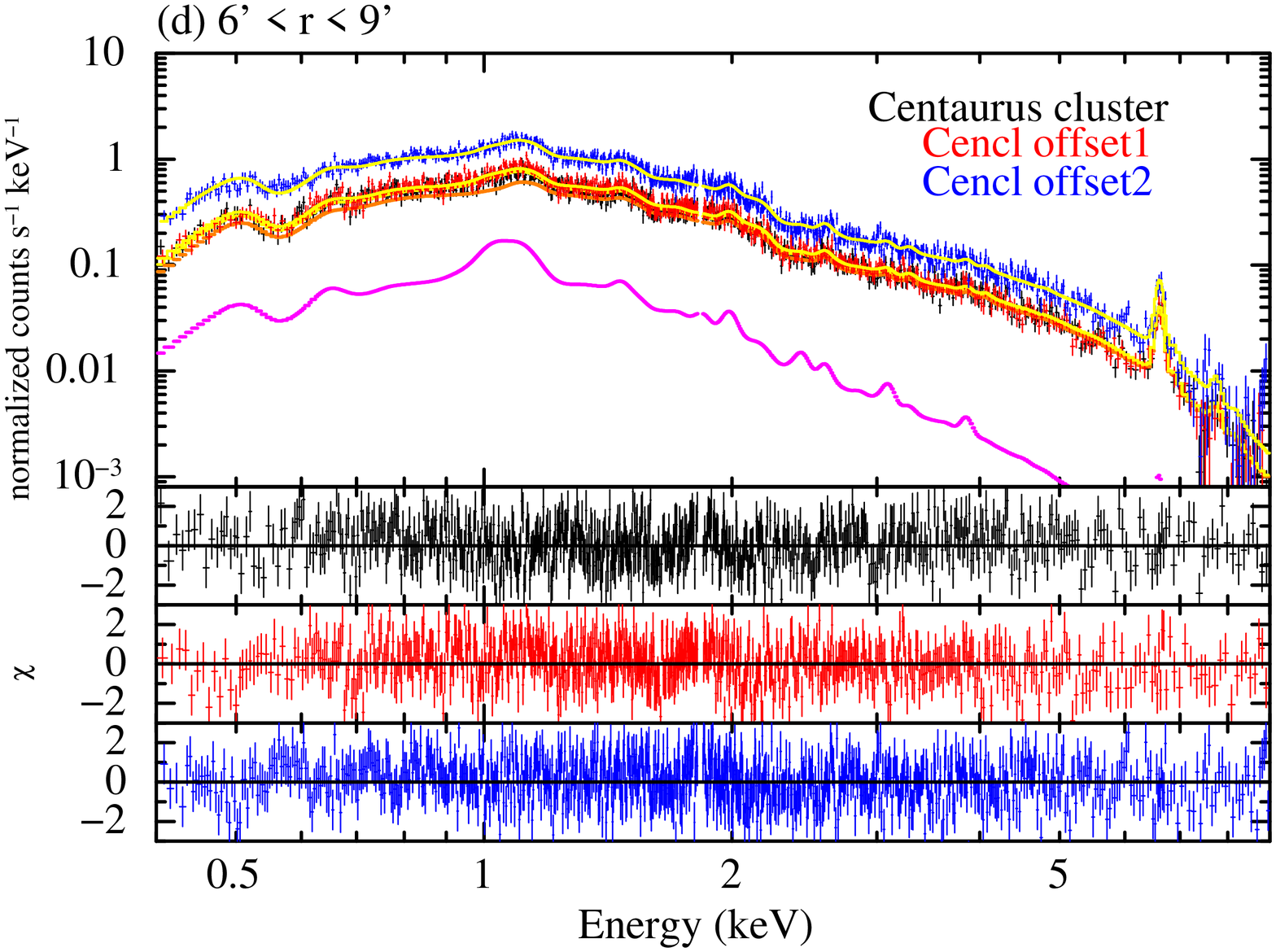}
\end{minipage}\hfill
\begin{minipage}{0.33\textwidth}
\FigureFile(55mm,55mm){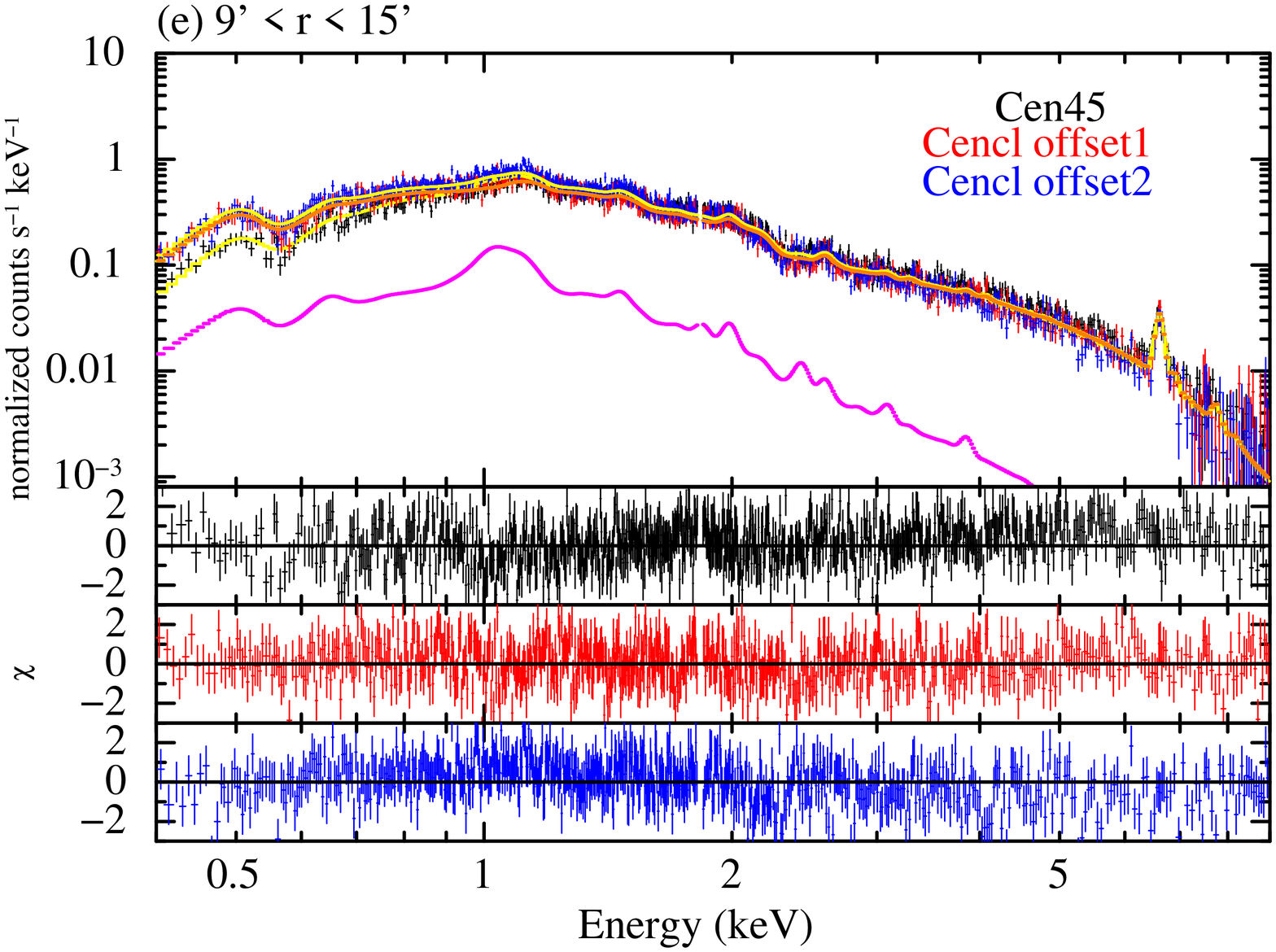}
\end{minipage}\hfill
\begin{minipage}{0.33\textwidth}
\FigureFile(55mm,55mm){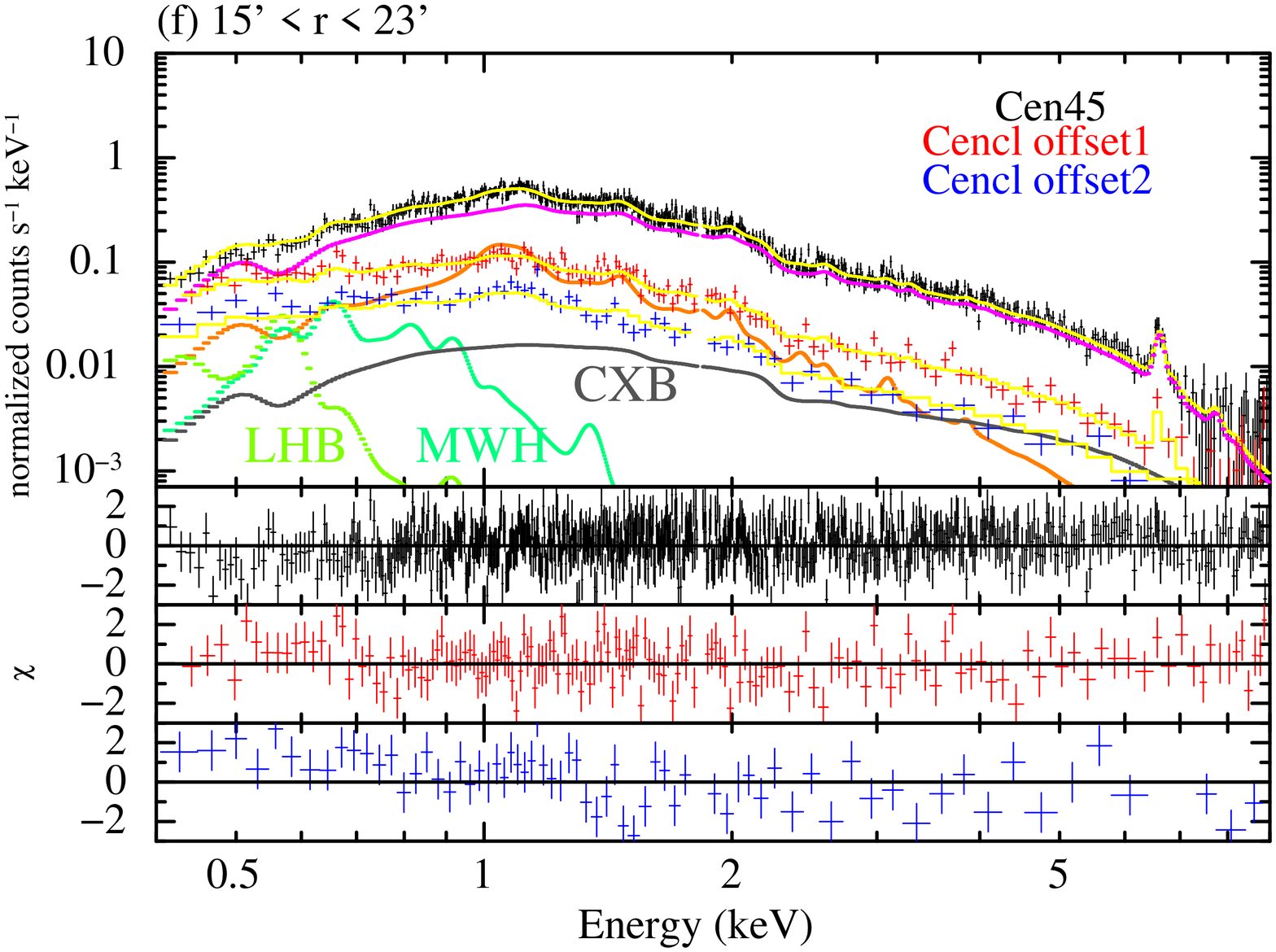}
\end{minipage}

\caption{
The panels show the observed spectra for the annular regions of 
the Centaurus cluster. The data 
are plotted by black, red, and blue crosses for XIS~1 spectra of 
the CENTAURUS\_CLUSTER, CENCL\_OFFSET1, and CENCL\_OFFSET2 region 
in 0$'$--9$'$, while CEN45, CENCL\_OFFSET1, and CENCL\_OFFSET2 
in 9$'$--23$'$, respectively. The estimated NXB, CXB, and 
Galactic components were subtracted from the spectra in the 0$'$-15$'$ 
region.  In 15$'$-23$'$, the CXB component is a fixed Lockmanhole value
 and Galactic components are included as the models.  
The yellow lines show the best-fit model 
for XIS~1.  The XIS~1 spectra of the ICM components are shown 
in magenta, orange, and light blue lines. The energy range around 
the Si--K edge (1.82--1.84 keV) was ignored in the spectral fits. 
The lower panels show the fit residuals in units of $\sigma$.}
\label{fig:fitresultimage}
\end{figure*}

\subsection{ICM Component}

We extracted spectra from six annular regions of 0$'$--2$'$, 2$'$--4$'$, 
4$'$--6$'$, 6$'$--9$'$, 9$'$--15$'$, and 15$'$--23$'$, centered on 
the peak of X-ray intensity at 
(RA, Dec)=(\timeform{192D2058}, \timeform{-41D3042}) \citep{Ota2007}.
The annular spectra are shown in figure 2.
Each spectrum was binned carefully to observe details in metal lines,
especially below $\sim1$~keV\@. Each spectral bin contained 50 or more counts.
We fitted the spectra in 0.5--9.0 keV at XIS~0, 2, 3 and 0.4--9.0 keV 
at XIS~1.  We excluded the narrow energy band around the Si--K edge 
(1.82--1.84 keV) because of an anomalous response. 

We assumed a single- or two-temperature (hereafter, 1T or 2T, respectively) 
model 
($vapec$; \cite{Smith2001}) and fitted the spectra of all detectors 
simultaneously for each region. 
We also examined a three-temperature 
(3T) model at the innermost region within $2'$, because the Chandra 
result suggested that at least a 3T model was needed 
to represent the spectra in the central region \citep{Sanders2006}.

The metal abundances of He, C, N, and Al were fixed to a solar value, 
while the other nine elements were allowed to vary. Note that Ne--K 
and Ni--L lines could not be resolved from the Fe--L lines. Therefore, 
we did not report Ne and Ni abundances in this paper.  The Galactic
absorption for neutral hydrogen $N_{\rm H}$ was also allowed to
vary in the spectral fits.
This is because the galactic latitude of the Centaurus, \timeform{21D7}, is relatively 
low, and there may be systematic uncertainties in the adopted Galactic absorption
which is a weighted average of those of positions within 1 degree from the cD galaxy.
In addition, a systematic uncertainty in the thickness of
the  contaminant on the optical filter of the XIS also causes a systematic uncertainty in the $N_{\rm H}$.

\section{Results}

\begin{figure}[t]
\begin{center}
\FigureFile(80mm,80mm){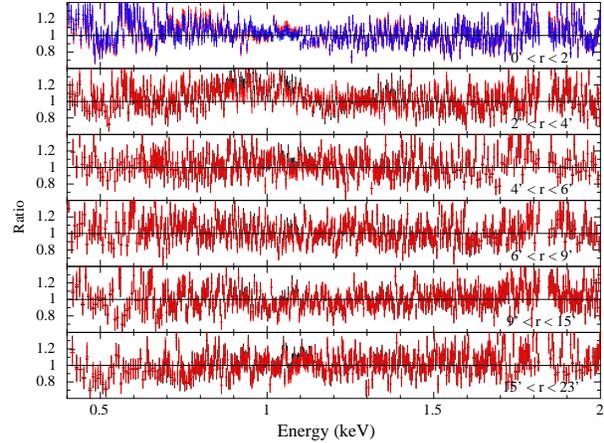}
\end{center}
\caption{The data-to-model ratios of the fits of the XIS~1 spectra
 with the 1T (black), 2T (red), and 3T (blue) models.}
\label{fig:residuals}
\end{figure}

\begin{figure*}[t]
\begin{minipage}{0.5\textwidth}
\FigureFile(80mm,80mm){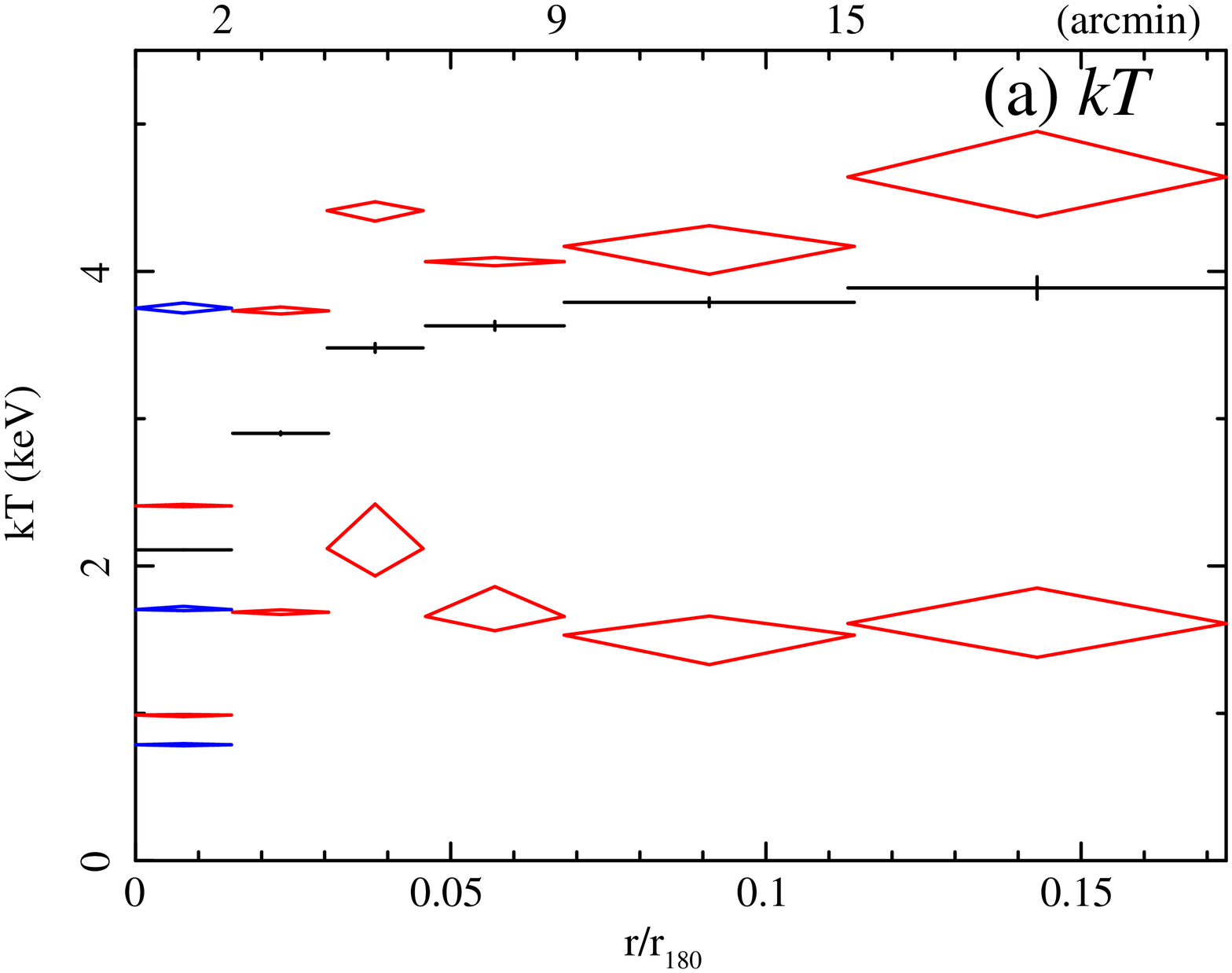}
\end{minipage}\hfill
\begin{minipage}{0.5\textwidth}
\FigureFile(80mm,80mm){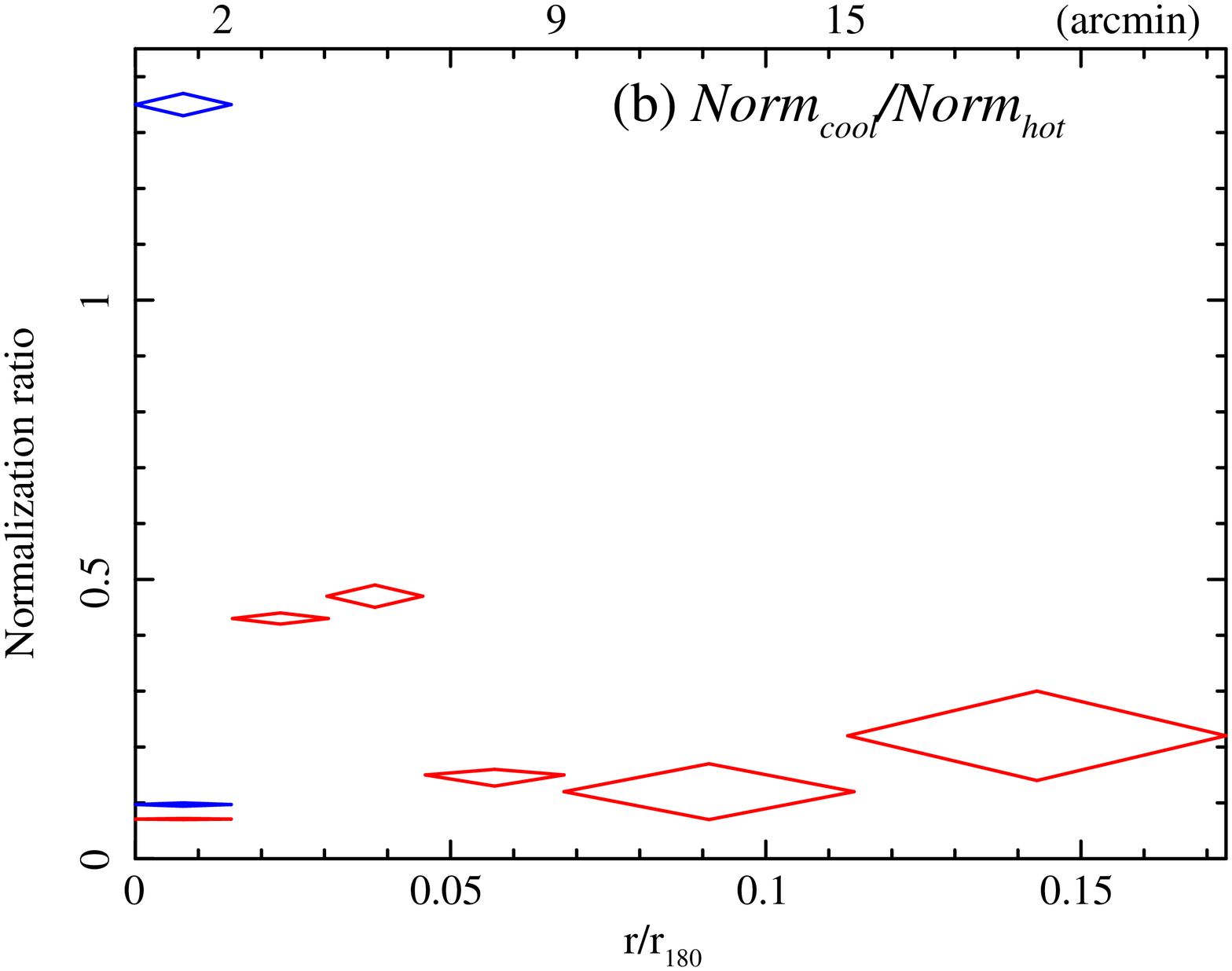}
\end{minipage}
\caption{
(a) Radial temperature profile derived from the spectral fits with 
the 1T (black cross), 2T (red diamond), and 3T (blue diamond) models. 
(b) Ratios of the normalization of the hotter to the cooler ICM 
component obtained from the spectral fits with the 2T (red diamond) and 
3T (blue diamond) models. 
}
\label{fig:tempradialresults}
\end{figure*}

\begin{table*}[t]
\caption{Summary of the resultant parameters of fits to 
each annular spectrum of the Centaurus cluster}
\label{tb:fitresultstable}
\begin{center}
\begin{tabular}{lllllllll}
\hline
$r$ & & $kT_{\rm hot}$ & $kT_{\rm cool1}$ & $kT_{\rm cool2}$ & 
$N_{\rm H}$ & $Norm_{\rm cool1}$ & $Norm_{\rm cool2}$ & 
$\chi^{2}$/d.o.f. \\
(arcmin) & & (keV) & (keV) & (keV) & (10$^{22}$cm$^{-2}$) 
& /$Norm_{\rm hot}$ & /$Norm_{\rm hot}$ & \\
\hline
 0$'$--2$'$ & 3T & 3.75$^{+0.04}_{-0.03}$ & 1.70$^{+0.02}_{-0.01}$ & 0.79$^{+0.01}_{-0.01}$ & 0.137$^{+0.002}_{-0.002}$ & 1.35$^{+0.02}_{-0.02}$ & 0.097$^{+0.003}_{-0.003}$ & 8589$\slash$7234 \\
 0$'$--2$'$ & 2T & 2.41$^{+0.01}_{-0.01}$ & 0.99$^{+0.01}_{-0.01}$ & & 0.132$^{+0.001}_{-0.002}$ & 0.071$^{+0.001}_{-0.001}$ & & 9405$\slash$7236 \\
 0$'$--2$'$ & 1T & 2.11 &  &  & 0.128 & & & 15904$\slash$7238 \\
 2$'$--4$'$ & 2T & 3.73$^{+0.03}_{-0.03}$ & 1.69$^{+0.02}_{-0.02}$ & & 0.127$^{+0.002}_{-0.001}$ & 0.43$^{+0.01}_{-0.01}$ & & 8505$\slash$7490 \\
 2$'$--4$'$ & 1T & 2.90$^{+0.01}_{-0.01}$ &  &  & 0.122$^{+0.002}_{-0.002}$ &  &  & 9370$\slash$7492 \\
 4$'$--6$'$ & 2T & 4.41$^{+0.06}_{-0.07}$ & 2.12$^{+0.30}_{-0.19}$ & & 0.118$^{+0.002}_{-0.002}$ & 0.47$^{+0.02}_{-0.02}$ & & 6794$\slash$6423 \\
 4$'$--6$'$ & 1T & 3.48$^{+0.03}_{-0.03}$ &  &  & 0.115$^{+0.002}_{-0.002}$ & & & 6981$\slash$6425 \\
 6$'$--9$'$ & 2T & 4.07$^{+0.03}_{-0.03}$ & 1.66$^{+0.20}_{-0.10}$ & & 0.120$^{+0.002}_{-0.002}$ & 0.15$^{+0.01}_{-0.02}$ & & 7475$\slash$6877 \\
 6$'$--9$'$ & 1T & 3.63$^{+0.03}_{-0.03}$ &  &  & 0.117$^{+0.002}_{-0.002}$ & & & 7587$\slash$6879 \\
 9$'$--15$'$ & 2T & 4.17$^{+0.14}_{-0.19}$ & 1.53$^{+0.13}_{-0.20}$ & & 0.109$^{+0.003}_{-0.003}$ & 0.12$^{+0.05}_{-0.05}$ & & 7470$\slash$6818 \\
 9$'$--15$'$ & 1T & 3.79$^{+0.03}_{-0.03}$ & & & 0.106$^{+0.002}_{-0.002}$ & & & 7562$\slash$6820 \\
 15$'$--23$'$ & 2T & 4.64$^{+0.31}_{-0.27}$ & 1.61$^{+0.24}_{-0.23}$& & 0.112$^{+0.013}_{-0.012}$ & 0.22$^{+0.08}_{-0.08}$ & & 2386$\slash$2254 \\
 15$'$--23$'$ & 1T & 3.89$^{+0.08}_{-0.08}$ & & & 0.099$^{+0.011}_{-0.010}$ & & & 2436$\slash$2256 \\
\hline\hline
$r$ &  & O & Mg & Si & S & Ar & Ca & Fe \\
(arcmin) & & (solar) & (solar) & (solar) & (solar) & (solar) & (solar) & (solar)\\
\hline
 0$'$--2$'$ & 3T & 1.00$^{+0.08}_{-0.08}$ & 1.02$^{+0.05}_{-0.05}$ & 1.46$^{+0.03}_{-0.03}$ & 1.55$^{+0.04}_{-0.04}$ & 1.74$^{+0.11}_{-0.11}$ & 2.17$^{+0.16}_{-0.16}$ & 1.83$^{+0.02}_{-0.02}$ \\
 0$'$--2$'$ & 2T & 1.14$^{+0.07}_{-0.07}$ & 1.04$^{+0.06}_{-0.05}$ & 1.53$^{+0.03}_{-0.03}$ & 1.49$^{+0.04}_{-0.04}$ & 1.40$^{+0.10}_{-0.10}$ & 1.78$^{+0.12}_{-0.13}$ & 2.09$^{+0.05}_{-0.01}$ \\
 0$'$--2$'$ & 1T & 0.38 & 0.43 & 0.94 & 0.94 & 0.85 & 1.58 & 1.45 \\
 2$'$--4$'$ & 2T & 0.83$^{+0.09}_{-0.08}$ & 0.66$^{+0.06}_{-0.06}$ & 1.01$^{+0.03}_{-0.03}$ & 1.08$^{+0.05}_{-0.05}$ & 1.25$^{+0.12}_{-0.12}$ & 1.56$^{+0.15}_{-0.15}$ & 1.35$^{+0.01}_{-0.02}$ \\
 2$'$--4$'$ & 1T & 0.71$^{+0.12}_{-0.12}$ & 0.72$^{+0.08}_{-0.08}$ & 1.06$^{+0.04}_{-0.04}$ & 1.00$^{+0.05}_{-0.05}$ & 0.94$^{+0.11}_{-0.11}$ & 1.32$^{+0.13}_{-0.13}$ & 1.56$^{+0.02}_{-0.02}$ \\
 4$'$--6$'$ & 2T & 0.74$^{+0.14}_{-0.12}$ & 0.57$^{+0.08}_{-0.08}$ & 0.77$^{+0.05}_{-0.05}$ & 0.65$^{+0.06}_{-0.06}$ & 0.86$^{+0.16}_{-0.16}$ & 1.02$^{+0.19}_{-0.18}$ & 0.90$^{+0.02}_{-0.02}$ \\
 4$'$--6$'$ & 1T & 0.77$^{+0.16}_{-0.16}$ & 0.71$^{+0.11}_{-0.10}$ & 0.82$^{+0.06}_{-0.06}$ & 0.67$^{+0.07}_{-0.07}$ & 0.67$^{+0.16}_{-0.16}$ & 0.89$^{+0.17}_{-0.17}$ & 0.98$^{+0.02}_{-0.02}$ \\
 6$'$--9$'$ & 2T & 0.60$^{+0.12}_{-0.12}$ & 0.45$^{+0.08}_{-0.08}$ & 0.44$^{+0.05}_{-0.05}$ & 0.47$^{+0.06}_{-0.06}$ & 0.71$^{+0.16}_{-0.16}$ & 0.73$^{+0.18}_{-0.18}$ & 0.64$^{+0.02}_{-0.02}$ \\
 6$'$--9$'$ & 1T & 0.56$^{+0.16}_{-0.16}$ & 0.52$^{+0.10}_{-0.10}$ & 0.44$^{+0.05}_{-0.05}$ & 0.44$^{+0.07}_{-0.07}$ & 0.56$^{+0.16}_{-0.16}$ & 0.66$^{+0.17}_{-0.17}$ & 0.70$^{+0.02}_{-0.02}$ \\
 9$'$--15$'$ & 2T & 0.47$^{+0.17}_{-0.17}$ & 0.48$^{+0.10}_{-0.10}$ & 0.42$^{+0.06}_{-0.06}$ & 0.52$^{+0.08}_{-0.08}$ & 0.48$^{+0.19}_{-0.19}$ & 0.67$^{+0.21}_{-0.21}$ & 0.56$^{+0.02}_{-0.02}$ \\
 9$'$--15$'$ & 1T & 0.38$^{+0.18}_{-0.18}$ & 0.54$^{+0.11}_{-0.11}$ & 0.42$^{+0.06}_{-0.06}$ & 0.48$^{+0.08}_{-0.08}$ & 0.33$^{+0.18}_{-0.18}$ & 0.62$^{+0.19}_{-0.19}$ & 0.61$^{+0.02}_{-0.02}$ \\
 15$'$--23$'$ & 2T & 0.38$^{+0.80}_{-0.38}$ & 0.72$^{+0.20}_{-0.19}$ & 0.44$^{+0.11}_{-0.10}$ & 0.42$^{+0.15}_{-0.15}$ & 1.00$^{+0.41}_{-0.40}$ & 0.49$^{+0.46}_{-0.46}$ & 0.50$^{+0.04}_{-0.04}$ \\
 15$'$--23$'$ & 1T & 0.20$^{+0.72}_{-0.20}$ & 0.89$^{+0.24}_{-0.23}$ & 0.46$^{+0.12}_{-0.12}$ & 0.36$^{+0.15}_{-0.15}$ & 0.76$^{+0.37}_{-0.36}$ & 0.44$^{+0.39}_{-0.39}$ & 0.57$^{+0.04}_{-0.04}$ \\
\hline
\end{tabular}
\end{center}
\end{table*}

\begin{figure*}[t]
\begin{minipage}{0.33\textwidth}
\FigureFile(55mm,55mm){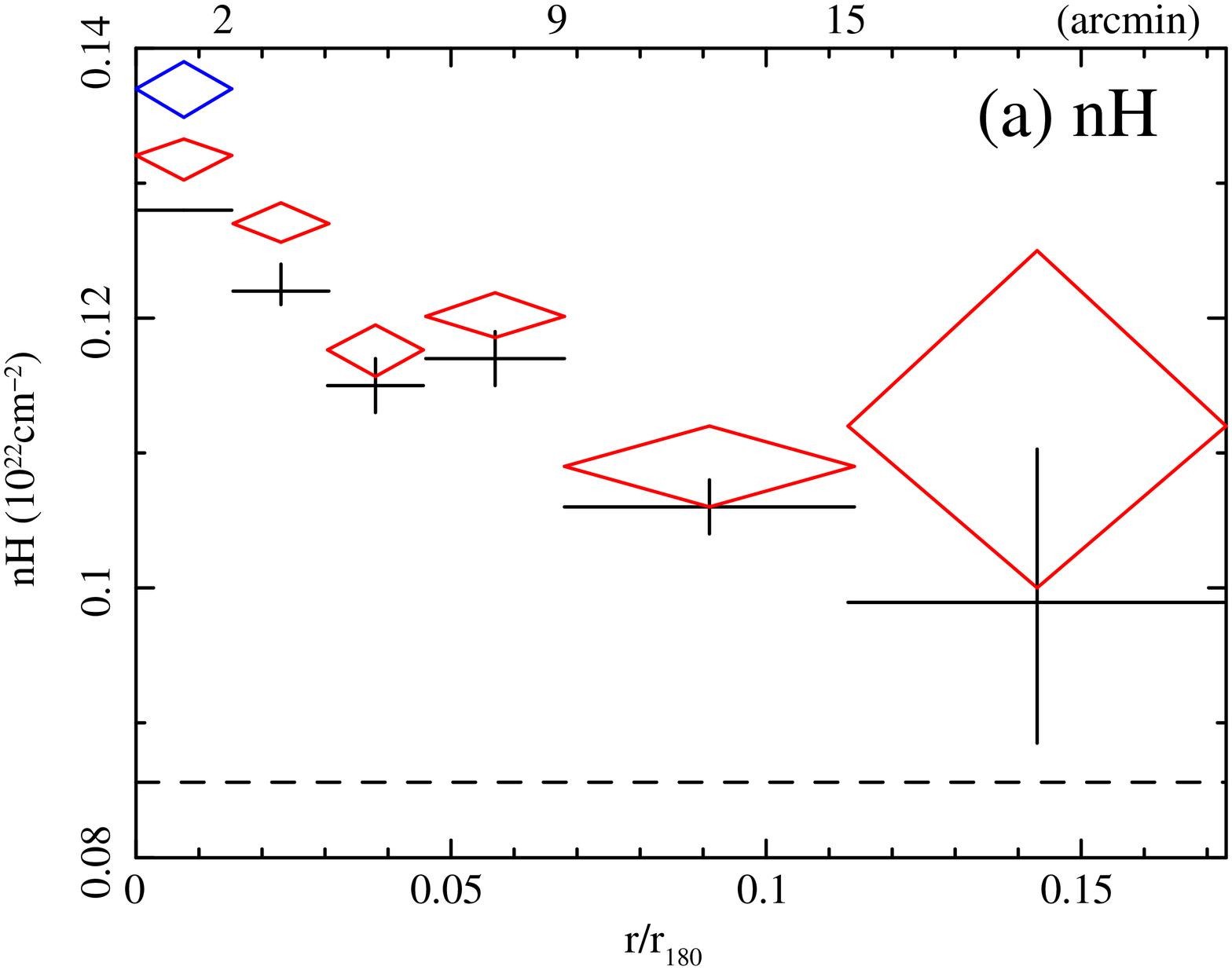}
\end{minipage}\hfill
\begin{minipage}{0.33\textwidth}
\FigureFile(55mm,55mm){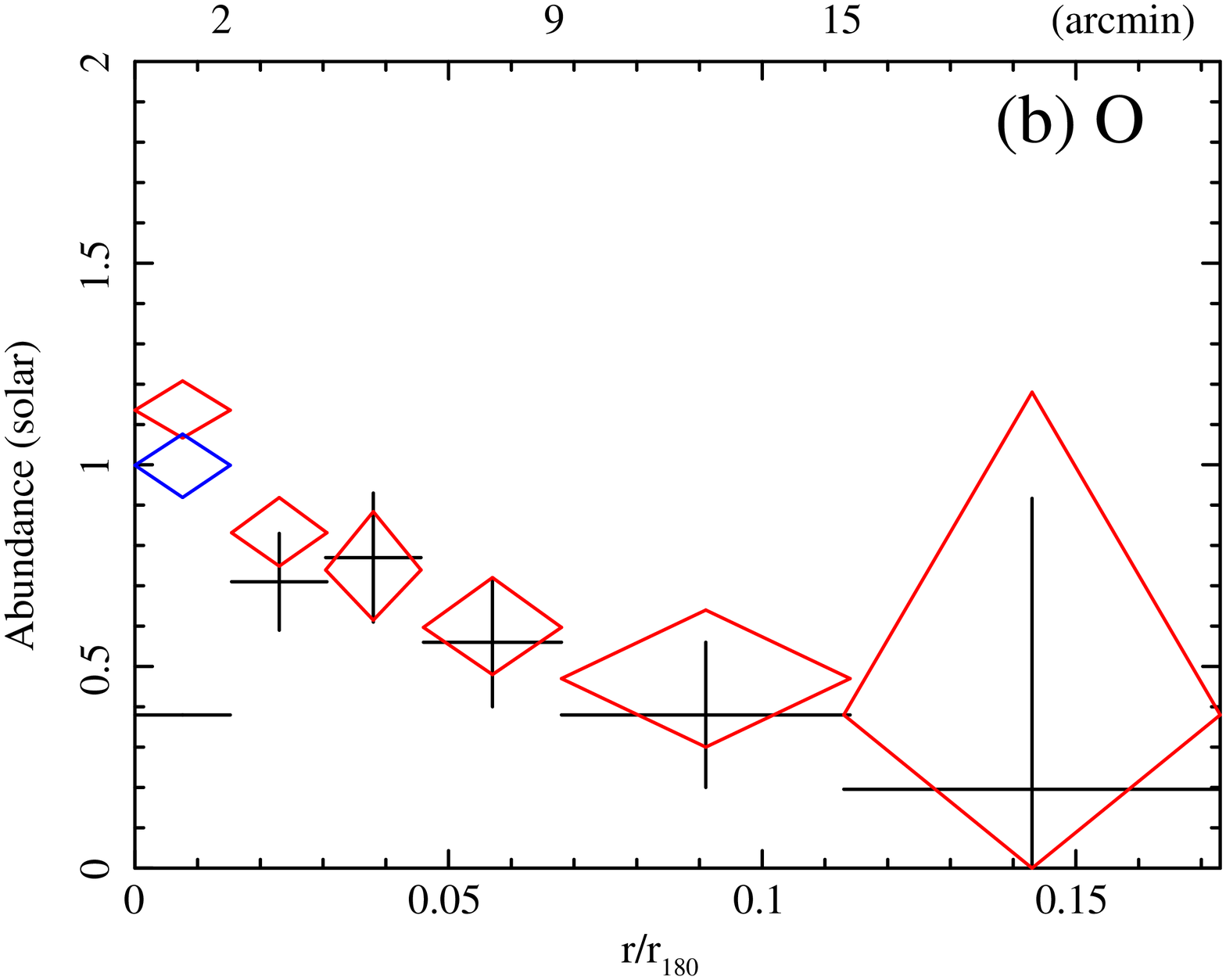}
\end{minipage}\hfill
\begin{minipage}{0.33\textwidth}
\FigureFile(55mm,55mm){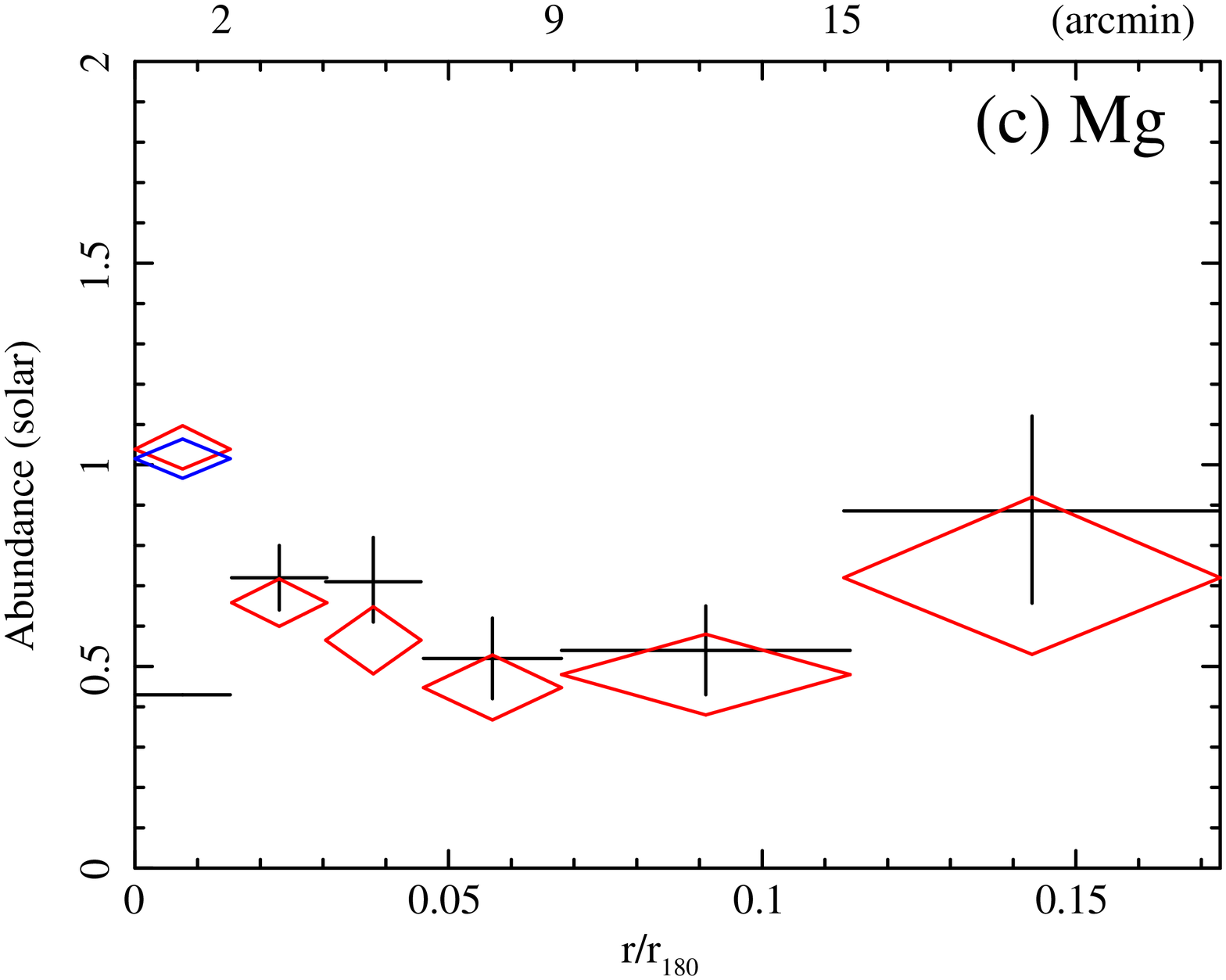}
\end{minipage}

\begin{minipage}{0.33\textwidth}
\FigureFile(55mm,55mm){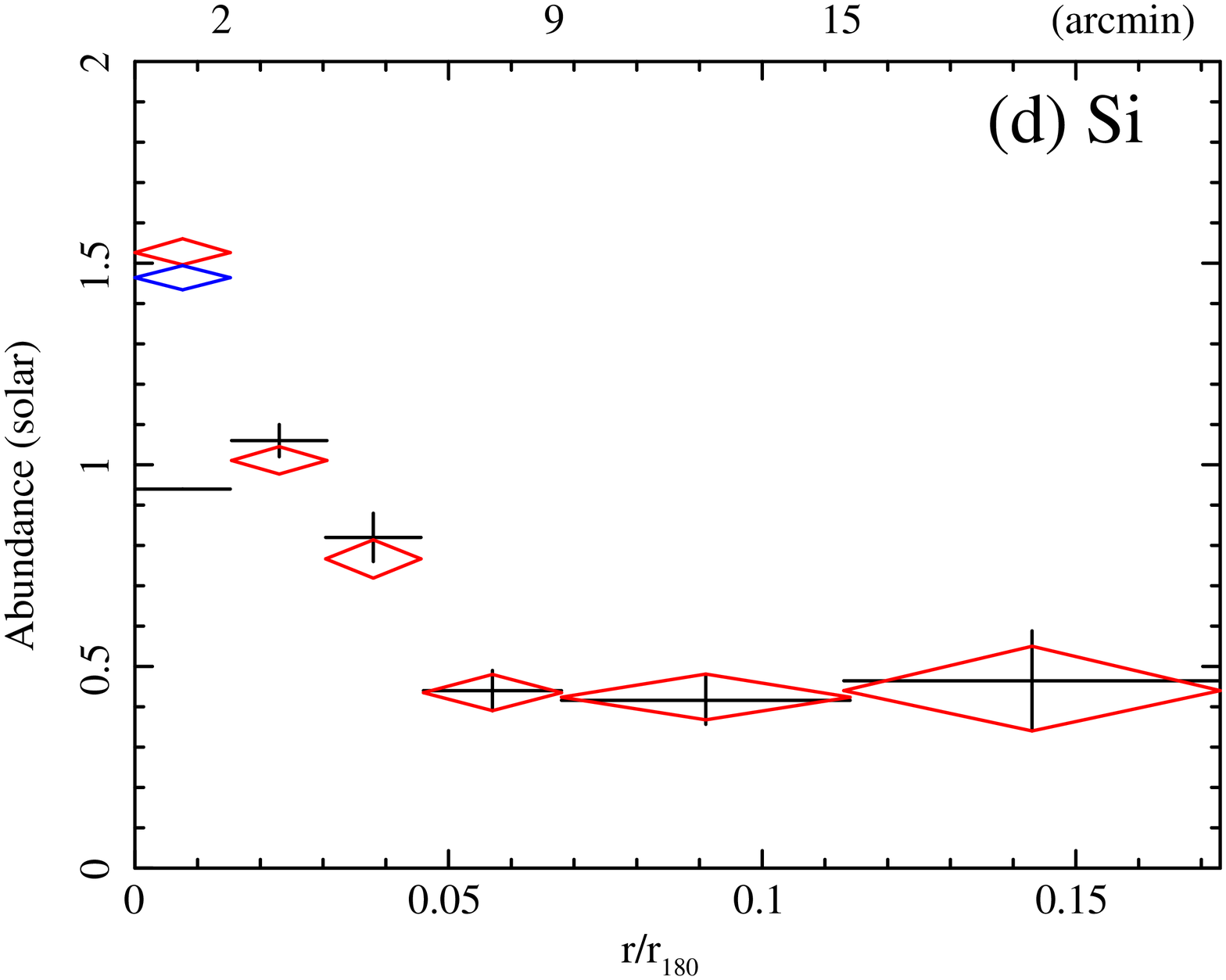}
\end{minipage}\hfill
\begin{minipage}{0.33\textwidth}
\FigureFile(55mm,55mm){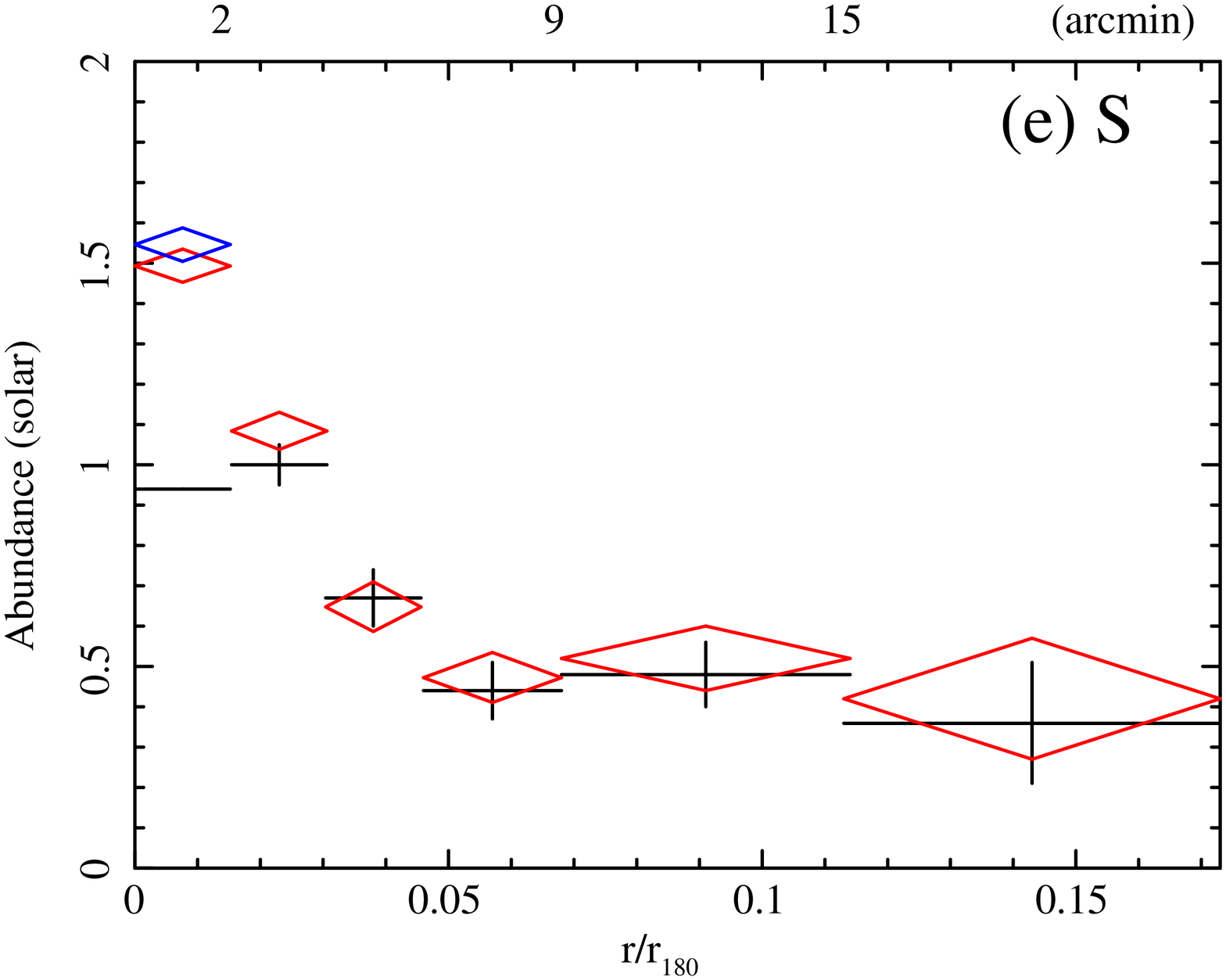}
\end{minipage}\hfill
\begin{minipage}{0.33\textwidth}
\FigureFile(55mm,55mm){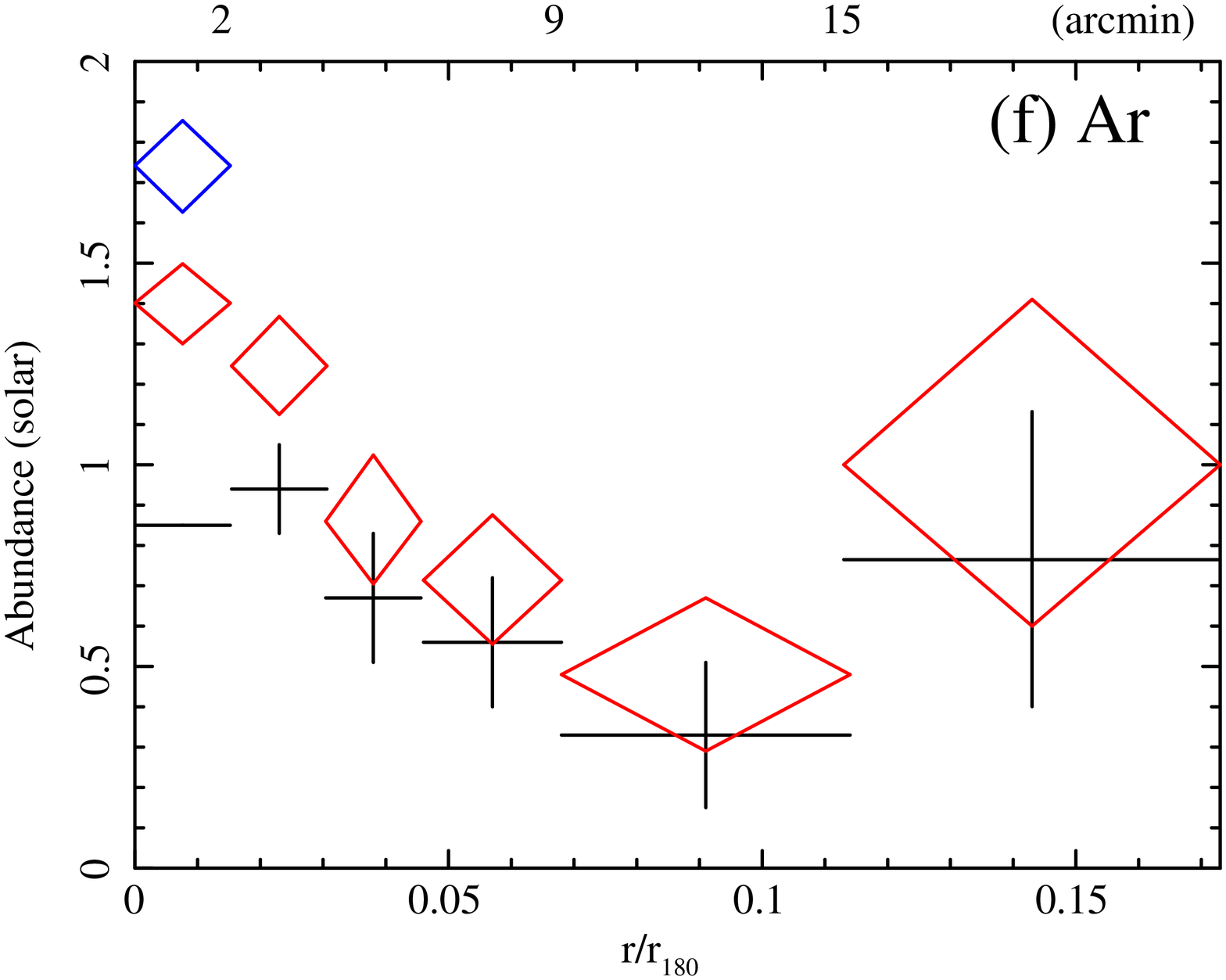}
\end{minipage}

\begin{minipage}{0.33\textwidth}
\FigureFile(55mm,55mm){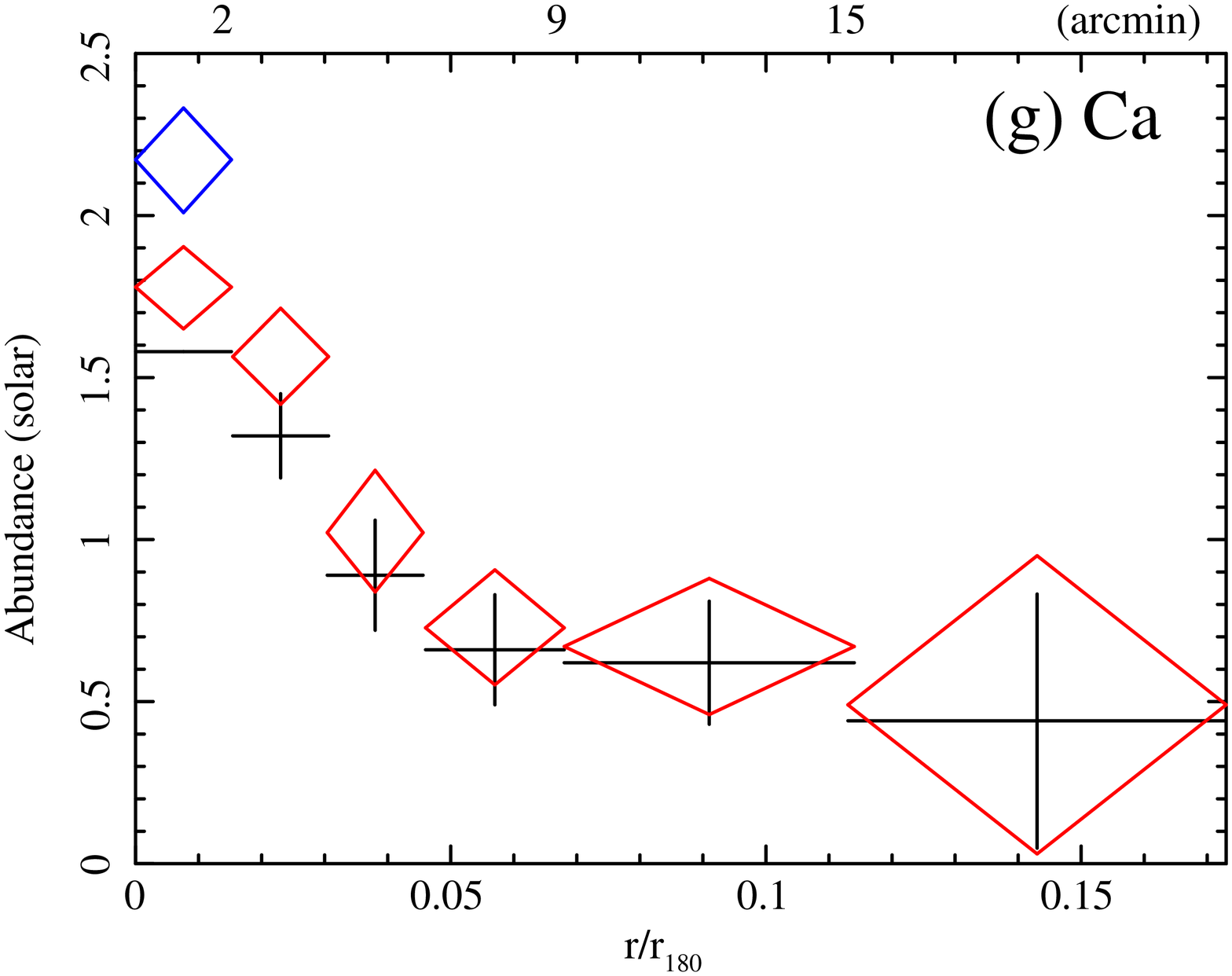}
\end{minipage}\hfill
\begin{minipage}{0.33\textwidth}
\FigureFile(55mm,55mm){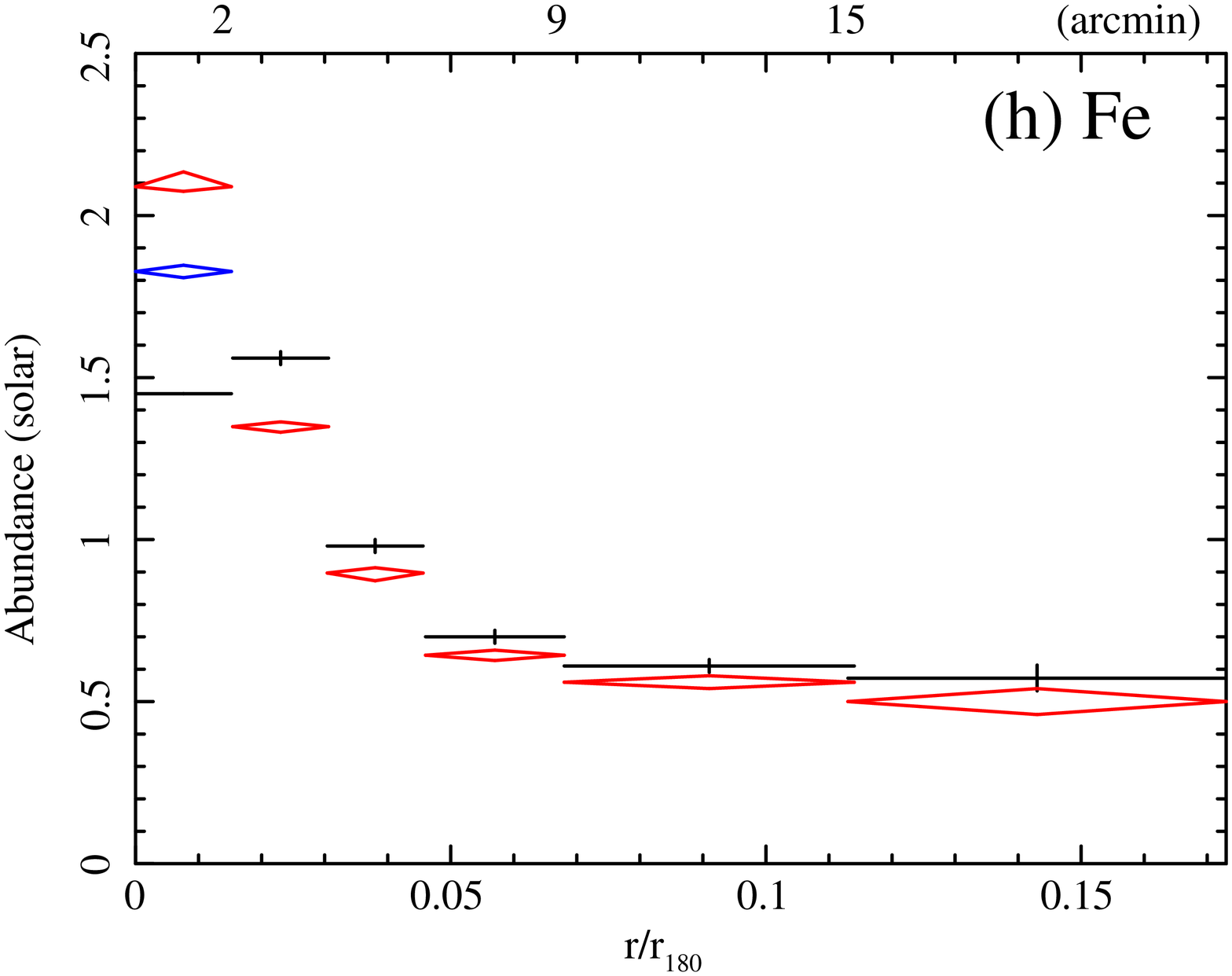}
\end{minipage}\hfill
\makebox[0.33\textwidth][l]{ }

\caption{
(a) Radial neutral hydrogen column density ($N_{\rm H}$) profile 
derived from the spectra. Black dashed lines show 
Galactic $N_{\rm H}$ of $8.56\times10^{20}$ cm$^{-2}$ 
from \citet{Kalberla2005}.  Black crosses, red and blue diamonds correspond 
to the results with the 1T, 2T, and 3T models, respectively. 
(b)--(h) Radial abundance profiles are plotted similar to that in (a).
}
\label{fig:abdradialresults}
\end{figure*}

\begin{figure*}[t]
\begin{minipage}{0.33\textwidth}
\FigureFile(55mm,55mm){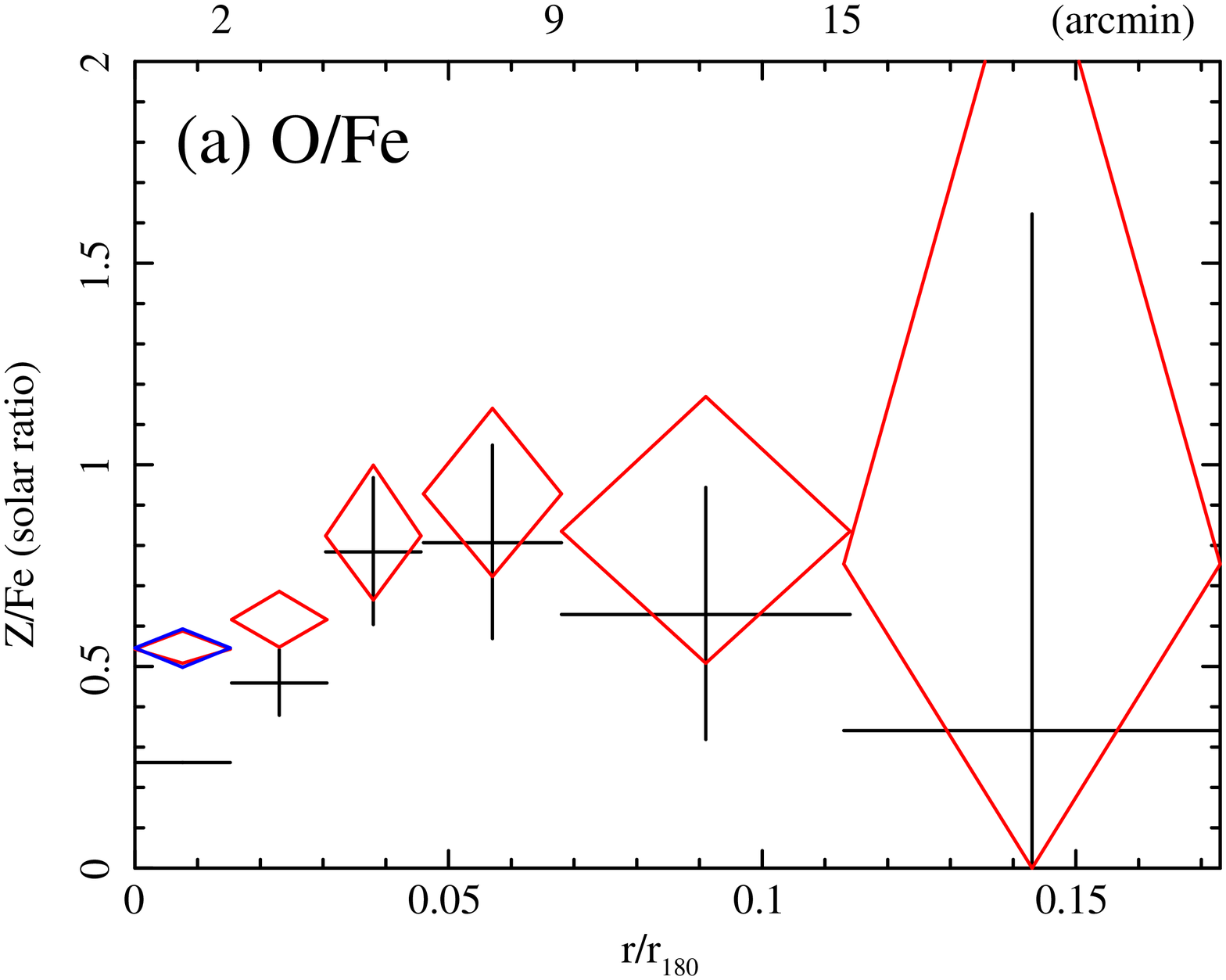}
\end{minipage}\hfill
\begin{minipage}{0.33\textwidth}
\FigureFile(55mm,55mm){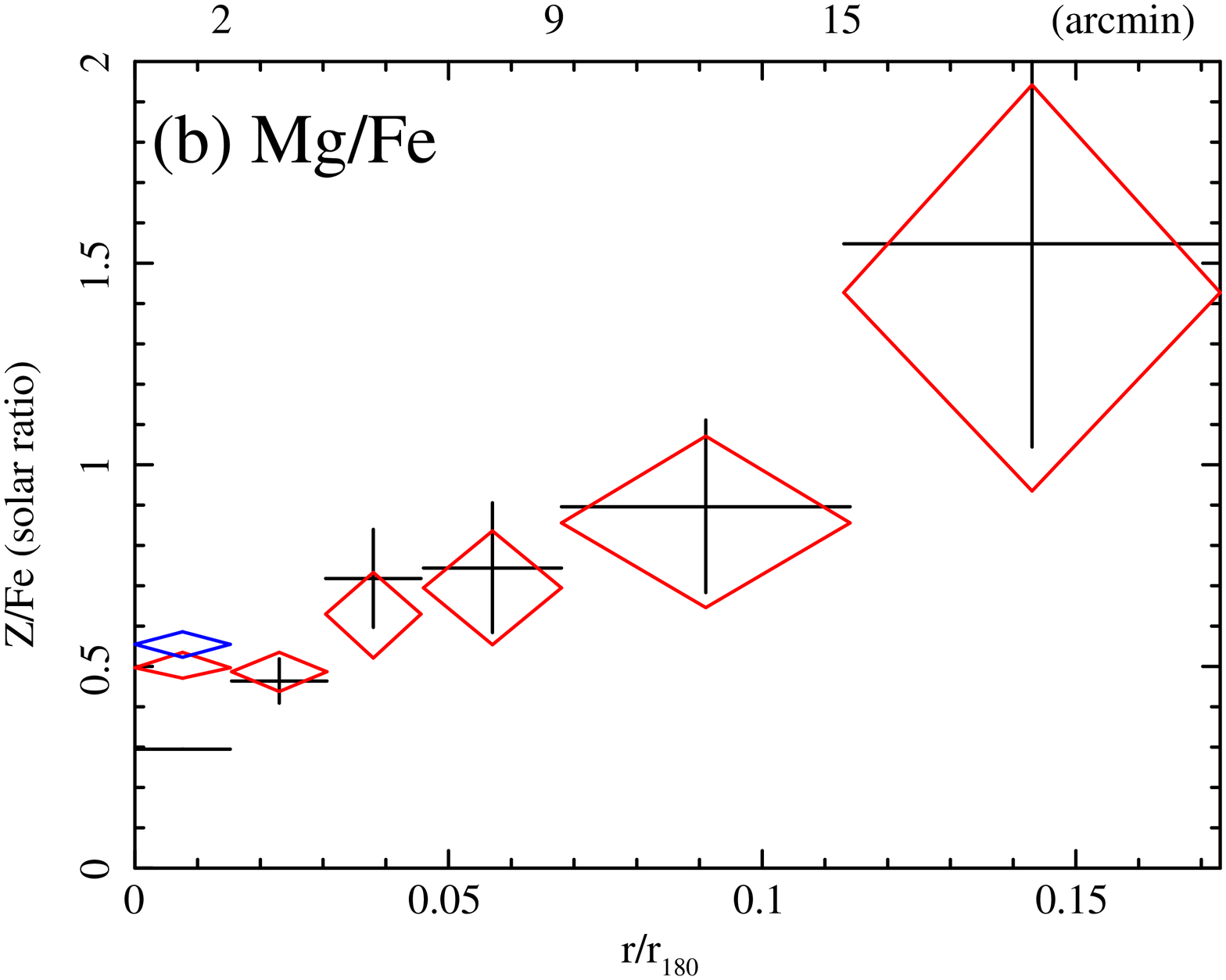}
\end{minipage}\hfill
\begin{minipage}{0.33\textwidth}
\FigureFile(55mm,55mm){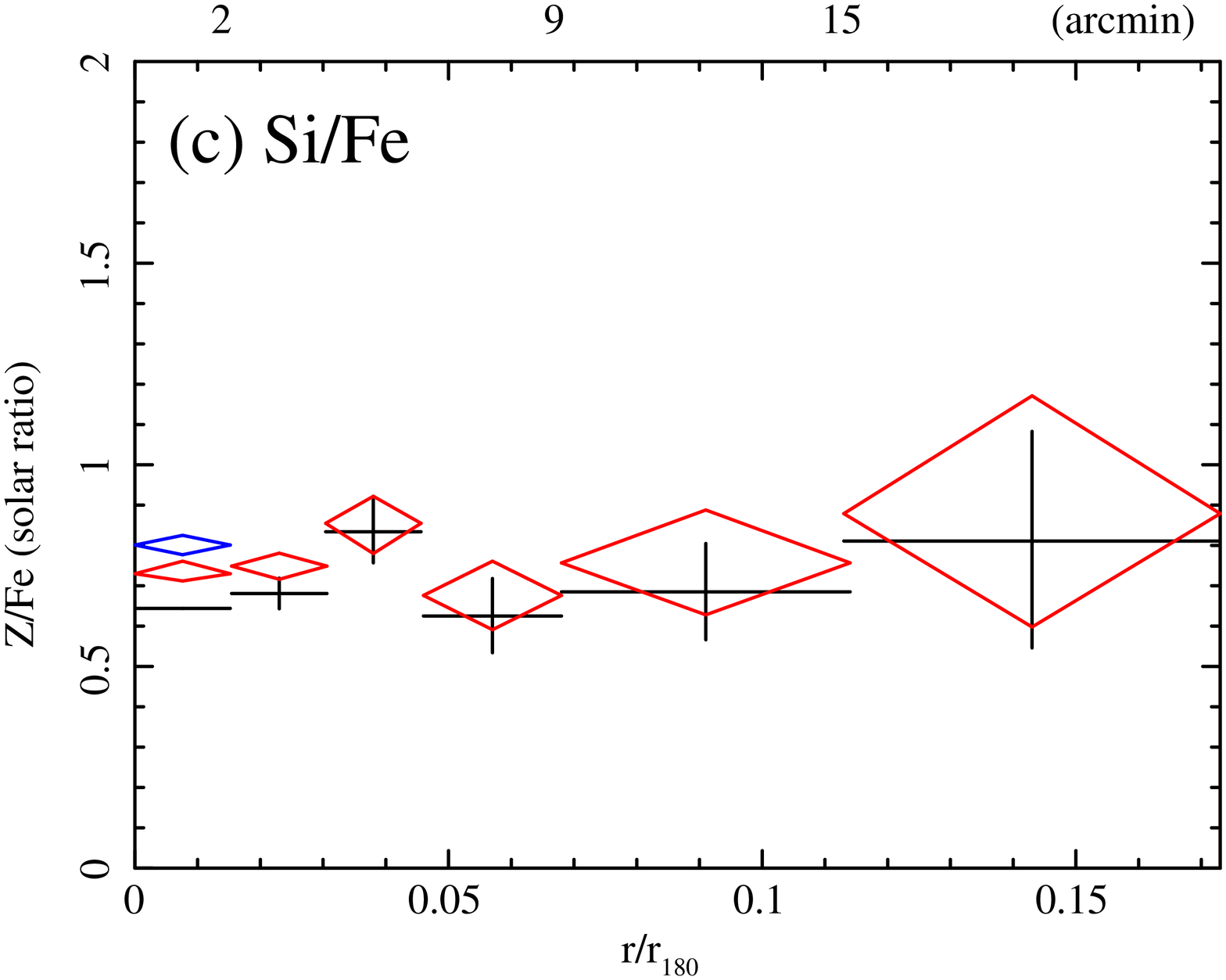}
\end{minipage}

\begin{minipage}{0.33\textwidth}
\FigureFile(55mm,55mm){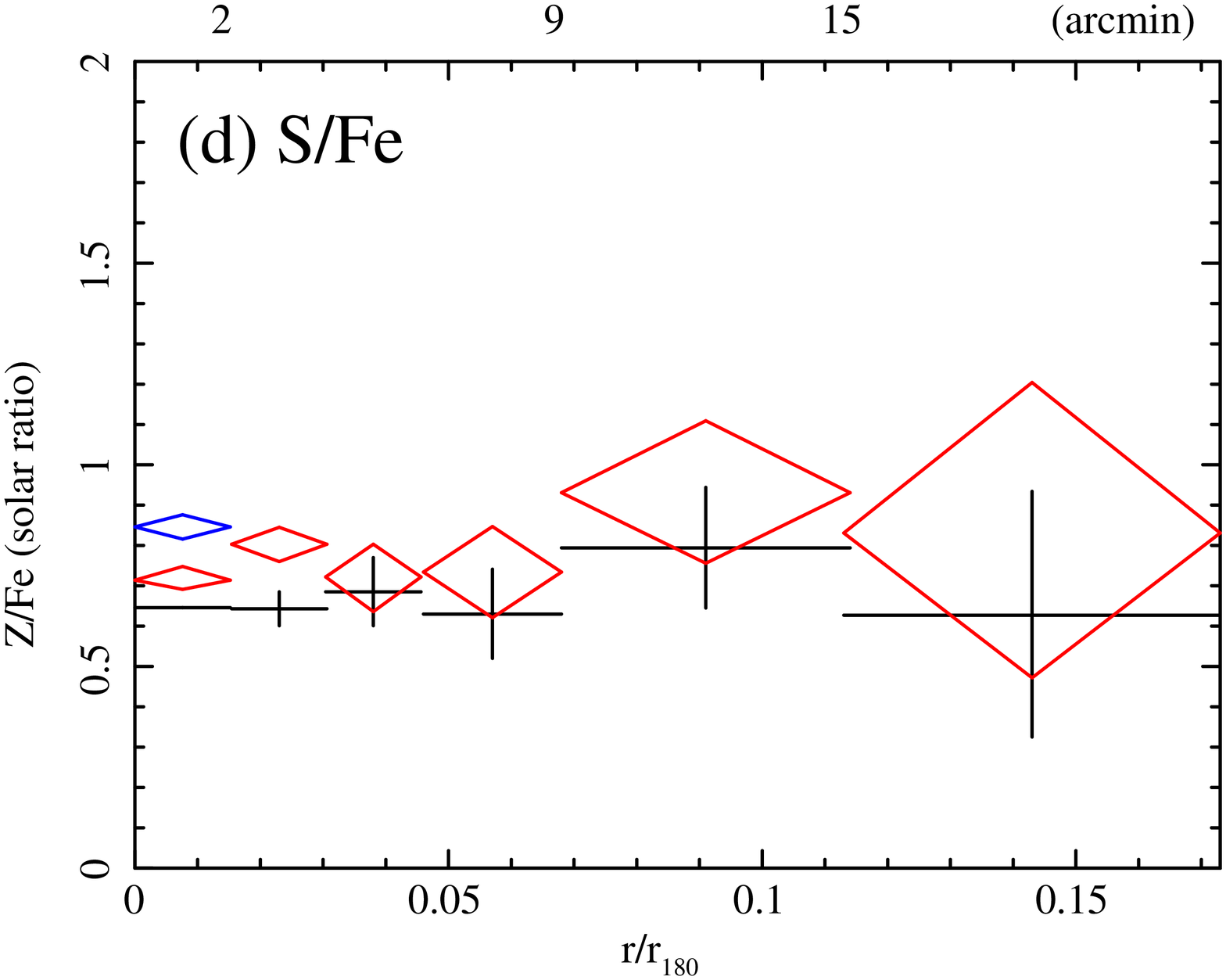}
\end{minipage}\hfill
\begin{minipage}{0.33\textwidth}
\FigureFile(55mm,55mm){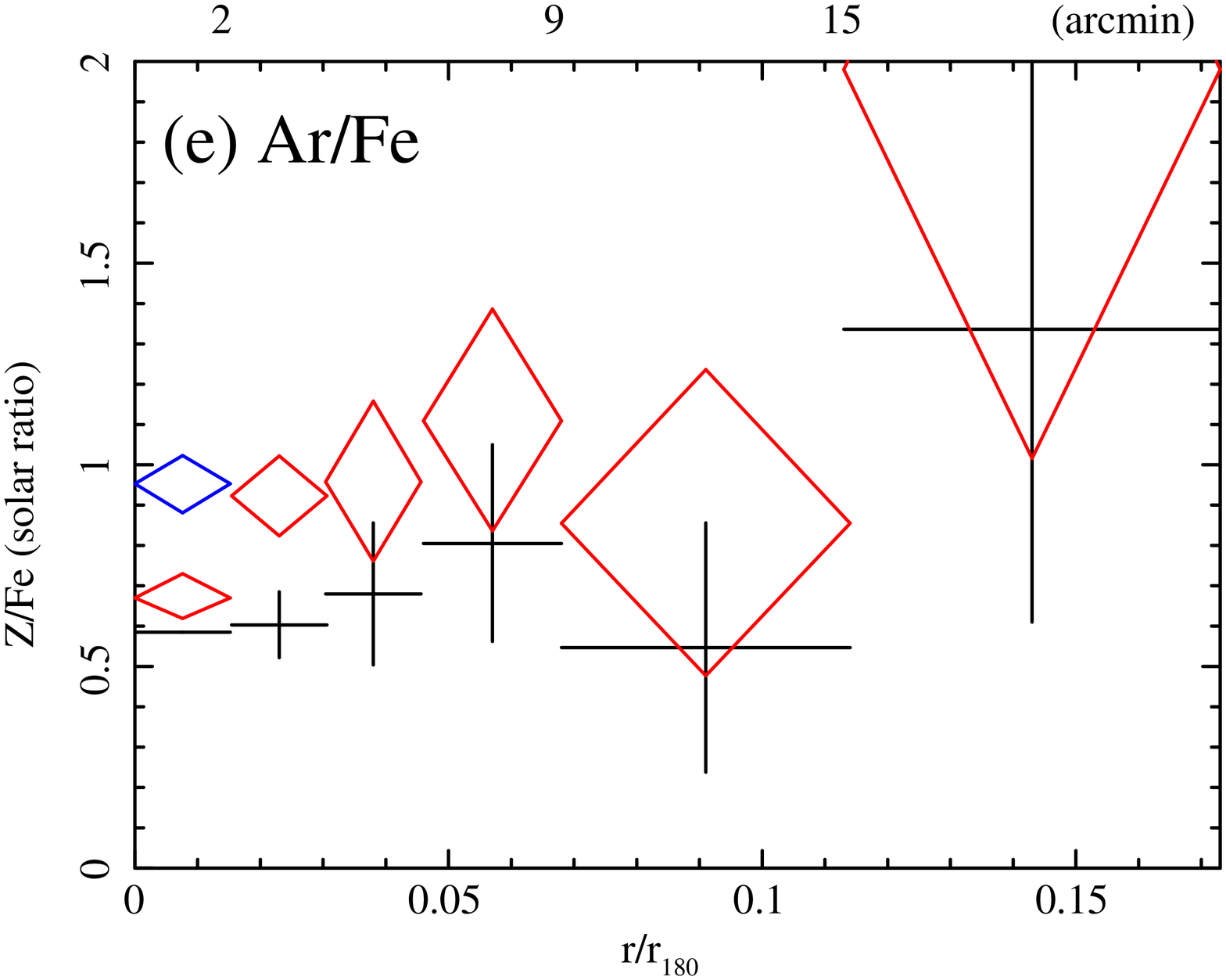}
\end{minipage}\hfill
\begin{minipage}{0.33\textwidth}
\FigureFile(55mm,55mm){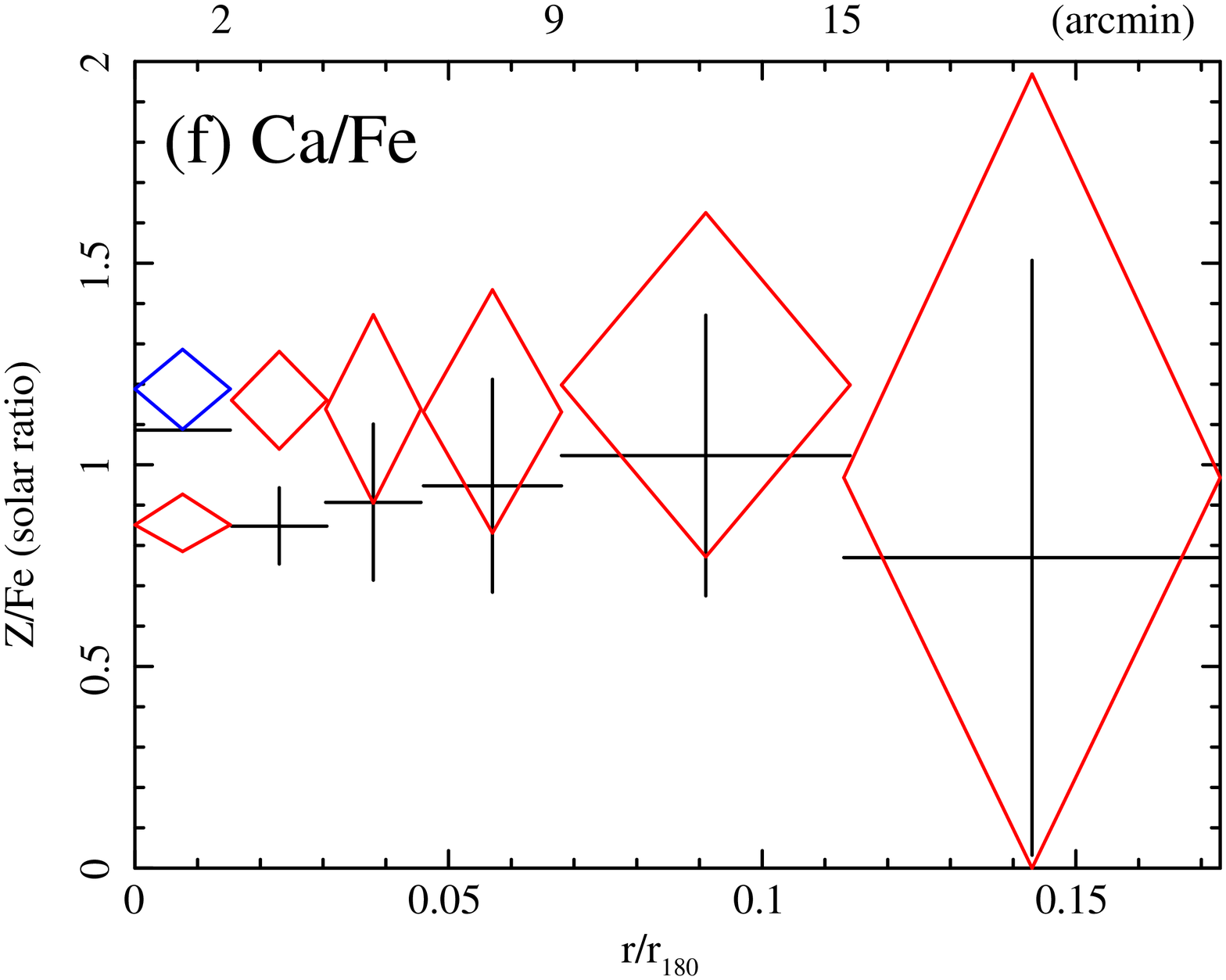}
\end{minipage}

\caption{Radial profiles of the abundance ratios of 
O, Mg, Si, S, and Ar to Fe in solar units \citep{Lodders2003}. 
The colors are the same as in figure \ref{fig:abdradialresults}.
}
\label{fig:abdratioradialresults}
\end{figure*}

\begin{figure*}[t]
\begin{minipage}{0.5\textwidth}
\FigureFile(80mm,80mm){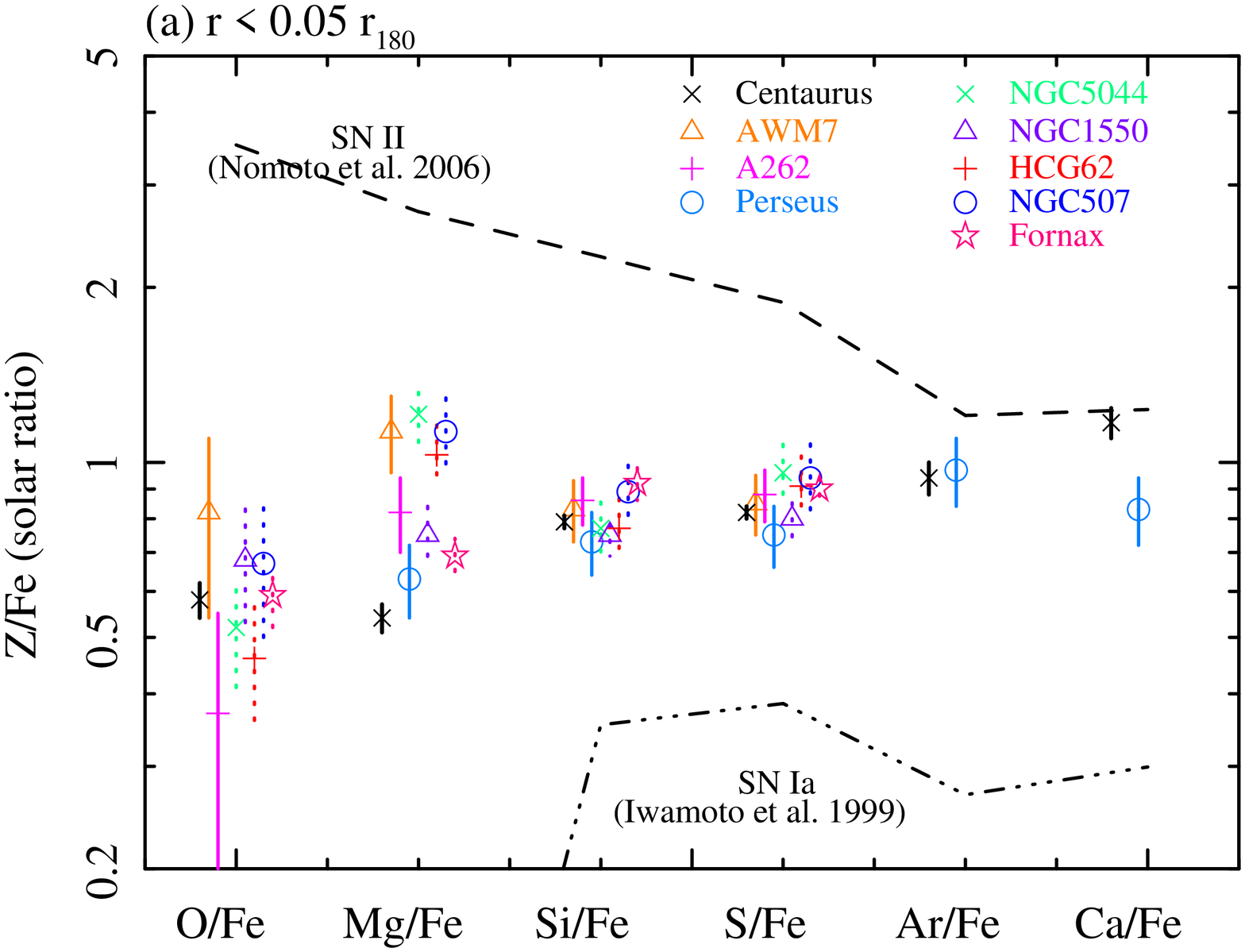}
\end{minipage}\hfill
\begin{minipage}{0.5\textwidth}
\FigureFile(80mm,80mm){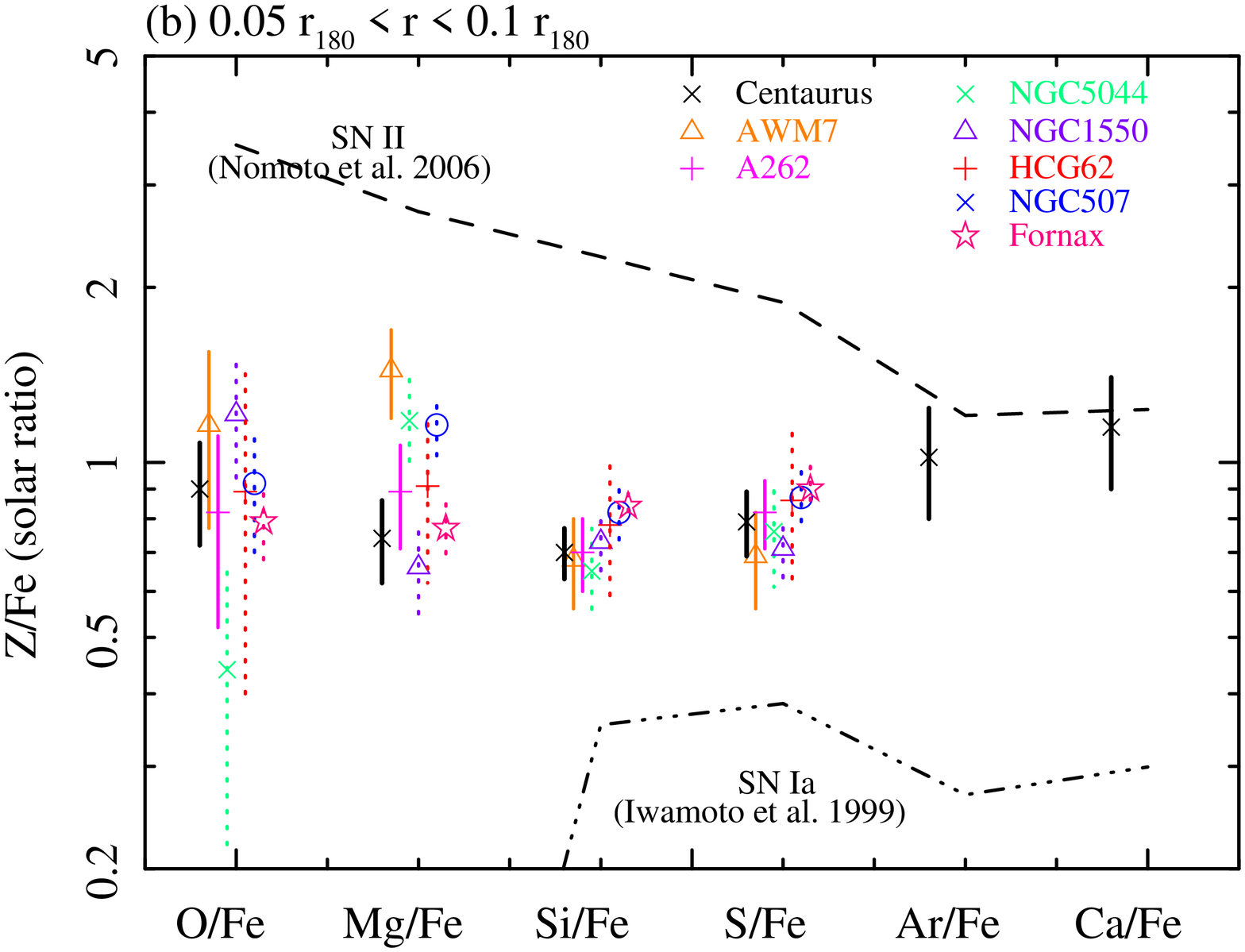}
\end{minipage}

\begin{minipage}{0.5\textwidth}
\FigureFile(80mm,80mm){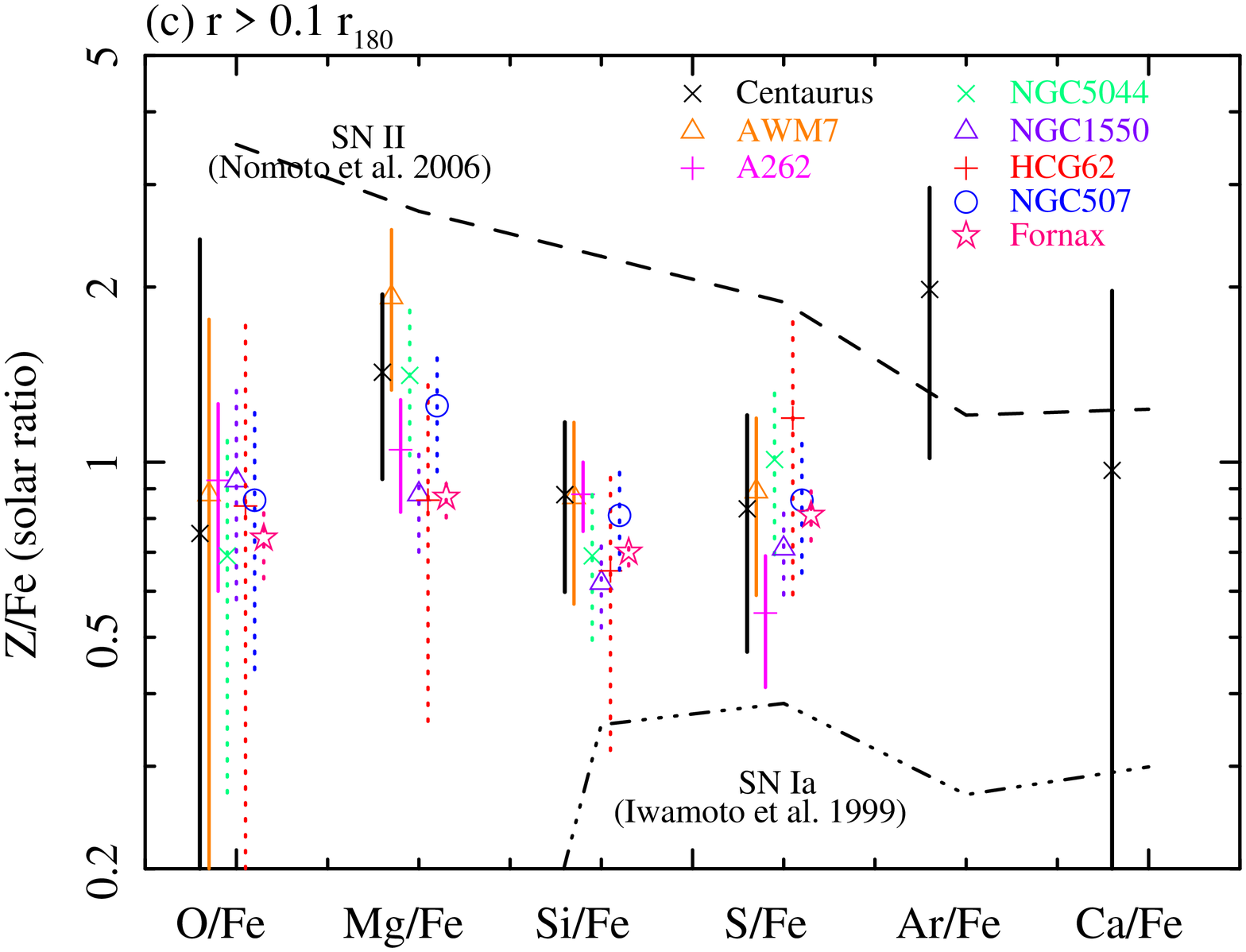}
\end{minipage}\hfill
\makebox[0.5\textwidth][l]{ }

\caption{Weighted averages of the abundance ratios of O, Mg, Si, S, 
Ar and Ca to Fe for the three regions (a) $r<0.05~r_{180}$, 
(b) $0.05<r<0.1~r_{180}$, and (c) $r>0.1~r_{180}$\@.  We used the 
abundance ratio derived from the 3T model within 0.015~$r_{180}$ 
and the 2T model for the outer regions. The abundance ratios of other 
clusters, AWM~7 \citep{Sato2008}, Abell~262 \citep{Sato2009b}, 
the Perseus cluster \citep{Tamura2009},
NGC~5044 \citep{Komiyama2009}, NGC~1550 \citep{Sato2010}, 
HCG~62 \citep{Tokoi2008}, NGC~507 \citep{Sato2009a}, and 
the Fornax cluster \citep{Matsushita2007a} are also plotted in the 
panels.
                                                                                 SN II yields by \citet{Nomoto2006} refer to an average of 
the Salpeter initial mass function of stellar masses 
from 10 to 50 $M\solar$ with a progenitor metallicity of $Z=0.02$.
The SN Ia yields were taken from the W7 model \citep{Iwamoto1999}.
}
\label{fig:abdratiocomp}
\end{figure*}

\subsection{Fitting Results}

Table~\ref{tb:fitresultstable} and figures~\ref{fig:residuals}, 
\ref{fig:tempradialresults}, 
and \ref{fig:abdradialresults} summarize the resultant parameters 
of the fits.  The data-to-model ratios of the fits are also shown in 
figure~\ref{fig:residuals}.  Within 2$'$, or $r<0.015~r_{180}$, 
the reduced $\chi^2$ with the 1T model is much larger than unity,
and the error ranges of parameters are not represented.
The 2T model significantly improved the resultant $\chi^2$ of the fit 
 compared to that with the 1T model.
  The fit with the  3T model is even better than 
that with the 2T model, particularly around 0.8 keV\@ (figure \ref{fig:residuals}). 
At 2$'$--4$'$, or $0.015$--$0.03~r_{180}$, 
the 2T model also improved the fits around 0.9--1.1 keV\@.
Beyond 4$'$, or $r>0.03~r_{180}$, 
with F-tests, the 2T model is significantly better than the 1T model,
considering only the statistical errors,
and  the data-model ratio improved by several percent around 1.1 keV.
However, considering possible systematic uncertainties in the Fe-L modeling,
the 1T model fits still represent the spectra fairly well.

\subsection{Temperature profile of the ICM}

The radial temperature profile and the ratios of the normalizations 
of the hotter and cooler ICM components from the cluster center to 
$r\sim0.17~r_{180}$ are shown in figure~\ref{fig:tempradialresults} 
and table~\ref{tb:fitresultstable}.  The ICM temperatures of the three 
components at the central region, within $0.015~r_{180}\simeq2'$, 
were $\sim3.8$, $\sim1.7$, and $\sim0.8$ keV, respectively, and 
the two higher temperatures with the 3T model appear to be similar 
to the hot and cool ICM temperatures with the 2T model in the 
$2'$--$4'$ ($0.015<r<0.03~r_{180}$) region.  Our results for the 
3T model are consistent with the previous Chandra result 
\citep{Sanders2006}.  For the outer region ($r>0.015~r_{180}$) 
the ICM temperature of the hotter component with the 2T model slightly
increased to $\sim4.6$ keV in the outermost region, while 
that of the cooler component was almost constant at $1.7$ keV\@.   
The resultant temperatures with the 1T model
were consistent with the previous XMM-Newton result with the 1T 
model \citep{Matsushita2007b}.  
A comparison of the normalization 
ratios of the hotter and cooler components for the 2T and 3T 
models is shown in figure~\ref{fig:tempradialresults}.  The ratios 
reached a peak at $\sim0.04~r_{180}$ and then decreased to 
approximately one fourth of its peak value in the outermost region.

\subsection{Abundance Profiles of ICM}

Radial abundance profiles of O, Mg, Si, S, Ar, Ca, and Fe up to 
$\sim0.17~r_{180}$ are summarized in table~\ref{tb:fitresultstable} 
and figure~\ref{fig:abdradialresults}. All the abundances lie 
over a solar abundance at the central region, and decrease toward 
the outer region. 

Within the cool core region ($r<0.045~r_{180}\simeq6'$) the 
abundances sharply increased toward the center.  
The Fe abundance 
reached a maximum value of $\sim$1.83 solar within 0.015~$r_{180}$ 
with the 3T model.  The Si, S, Ar, and Ca abundances showed the same 
behavior as the Fe profile.  The O and Mg abundances were approximately 1 
solar within 0.015~$r_{180}$.  

For the outer region ($r>0.045r_{180}$), 
the 1T and 2T model gave similar O, Mg, Si, S, Ca, and Fe abundances,
although the 2T model gave slightly higher abundances of Ar.
All the abundances were
nearly constant at $\sim$0.5 solar.  
Fe decreased to approximately one fourth of the central value in the outer region.  The ratio of O and Mg to Fe also 
decreased to approximately half of the central value. 
In the outermost region 
(0.11$<r<$0.17~$r_{180}$),  the fits gave only the upper limit of 
the O abundance with 90\% error.
It is the first time to derive Mg abundance to 0.17$r_{180}$.
The derived abundances are consistent with those from XMM and Chandra data
\citep{Sanders2006,Matsushita2007b}, but the error bars
are small particularly outside the cool core.

\subsection{Profiles of ICM Abundance Ratios}

The radial profiles of the abundance ratios of O, Mg, Si, S, Ar, and Ca to Fe 
abundance are shown in figure \ref{fig:abdratioradialresults}.
Excluding the central region, the 1T and 2T models gave similar abundance ratios of  O/Fe, Mg/Fe, Si/Fe, and S/Fe,
whereas the 2T model yielded higher Ar/Fe and Ca/Fe 
ratios by several tens of percent.
The abundance ratios of O$\slash$Fe and Mg$\slash$Fe 
are $\sim$0.5 solar ratio within 0.03 $r_{180}$ 
and  $\sim$0.7--1 solar ratio beyond the radius. 
The other abundance ratios of Si$\slash$Fe, S$\slash$Fe, 
Ar$\slash$Fe, and Ca$\slash$Fe are 
$\sim$1 solar ratio in the entire region.
The weighted averages of the abundance ratios are calculated for three radial regions
and are shown in table \ref{tb:abdratiotable}.

\subsection{Uncertainties}
\label{subsec:uncertainties}

In order to estimate systematic errors, we varied the normalizations 
of the NXB spectra and the mock spectra including the CXB, LHB, and 
MWH by $\pm$10\% in the spectral fits.  The systematic errors due to
 background estimation were almost negligible, and the resultant 
temperatures and abundances did not change within the statistical 
errors by changing the background level.  We also note that Ne 
abundance is not reliably determined because of an overlap with the strong 
and complex Fe--L line emissions. However, these abundances were
allowed  to vary freely during the spectral fits.

\section{Discussion}

\subsection{ICM Abundance Pattern and Contributions from SN Ia and SN II}

Figure \ref{fig:abdratiocomp} summarizes the abundance pattern of the O/Fe, Mg/Fe, Si/Fe, S/Fe, Ar/Fe, and Ca/Fe
within 0.05 $r_{180}$, 0.05--0.1 $r_{180}$, and beyond 0.1 $r_{180}$ of the
Centaurus, AWM 7, Abell~262, and the Perseus clusters with those of the four groups of galaxies and the Fornax cluster.
Within 0.1 $r_{180}$,
the Centaurus cluster has the smallest error bars for O/Fe and Mg/Fe ratios 
compared to the other clusters and groups of galaxies.

Outside the cool core, beyond 0.05 $r_{180}$, 
the O/Fe, Mg/Fe, Si/Fe, and  S/Fe ratios of these groups and clusters of galaxies are close to the solar ratio. 
The scatter in the Mg/Fe ratio might be caused by uncertainties in the Fe--L atomic data, because the Mg--K lines are surrounded by the Fe--L lines.
The Centaurus cluster and the Perseus cluster have similar Ar/Fe and Ca/Fe ratios.
There is no significant difference in the abundance pattern between the  clusters
and groups of galaxies.

Within 0.05 $r_{180}$, or the cool core region, 
Mg/Fe ratio of the Centaurus cluster shows the lowest value of 0.5 
in units of the solar ratio. Mg/Fe ratio 
of the other systems scatters around unity, i.e., the solar ratio.
 O/Fe ratio of the Centaurus cluster is the same as those of other clusters 
and groups of galaxies, although error bars of the other clusters and groups 
are fairly large.
In contrast, 
the Si/Fe and S/Fe ratios of the Centaurus clusters and other systems agree well.

\subsection{Mass-to-Light Ratios}
 The metal MLR is a useful measure for the study of ICM chemical evolution because metals in the ICM are all synthesized in galaxies.
 We calculated B-band and K-band luminosity profiles to estimate the MLRs.
The K-band luminosity of a galaxy correlates more than 
the B-band with stellar mass.
The B-band luminosities were used to compare previous studies.

To derive a B-band luminosity profile, 
we integrated the luminosities of the member galaxies 
from \citet{Jerjen1997}, deprojected the profile assuming spherical symmetry, 
and derived a three-dimensional profile of  $L_{\rm B}$.
\citet{Jerjen1997} optically studied the Centaurus cluster 
and identified 296 member galaxies whose apparent magnitudes 
are brighter than 21.5 in the B-band.
The brightest cluster galaxy NGC~4696 has an apparent magnitude 
$m_{\rm B}$ = 11.3 or log$L_{\rm B} \slash L_{\rm B,\solar}$ = 11.2 
using 44.9~Mpc and the foreground Galactic extinction $A_{\rm B}$ = 0.492 
\citep{Schlegel1998} from NASA$\slash$IPAC Extragalactic Database (NED).

\begin{figure}[t]
\begin{center}
\FigureFile(80mm,80mm){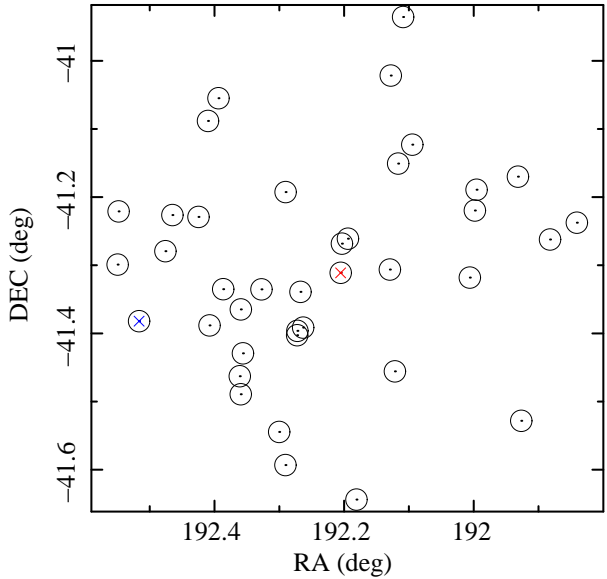}
\caption{Identified objects from the Two Micron All Sky Survey 
(2MASS) in K-band (black open circle). 
the red and blue crosses 
denote the coordinate of the cD galaxy (NGC~4696, \cite{Ota2007}) 
and NGC~4709, respectively.
}
\label{fig:Lkmem}
\end{center}
\end{figure}

To derive a K-band luminosity profile,
we used the luminosity of galaxy data in a box of 10$\times$10 deg$^{2}$ centered at the peak of X-ray intensity from the Two Micron All Sky Survey (2MASS) \footnote{http://www.ipac.caltech.edu/2mass/}. 
The brightest cluster galaxy NGC 4696 has an apparent magnitude $m_{\rm K}$ = 7.298 
or log $L_{\rm K}/L_{\rm K, \solar}$ = 11.7 using the foreground Galactic extinction $A_{\rm K}$ = 0.042 (Schlegel et al. 1998) from NED. 
Figure \ref{fig:Lkmem} shows the galaxies detected with 2MASS within $\sim$0.2 $r_{180}$
from the cD galaxy.
The average surface brightness in the region of $140'<r<300'$ (r$_{180} \sim140'$) was subtracted as the background.
Then, we calculated the deprojected three-dimensional profile of $L_{\rm K}$
from the brightness profile of $L_{\rm K}$.

\begin{figure}[t]
\begin{center}
\FigureFile(80mm,80mm){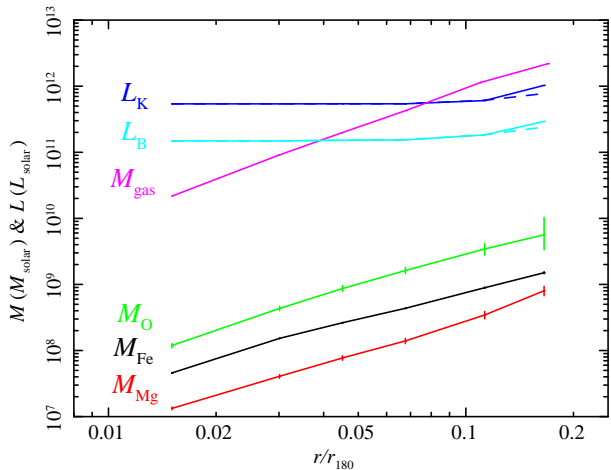}
\caption{
Integrated mass of the hot gas (magenta), O (green), 
Mg (red) and Fe (black).  
Integrated K-band luminosities of galaxies $L_{\rm K}$ (blue) and 
B-band luminosities of galaxies $L_{\rm K}$ (light blue)
 are also plotted in the panel.
Integrated luminosities excluding NGC~4709 are shown with dashed lines.
}
\label{fig:IntegratedMandL}
\end{center}
\end{figure}

Figure \ref{fig:IntegratedMandL} shows the integrated radial profiles of $L_{\rm B}$ and $L_{\rm K}$.
Within 0.1~$r_{180}$, the cD galaxy dominates the $L_{\rm B}$ and $L_{\rm K}$ profiles.
Uncertainties in the three-dimensional position of bright galaxies except the cD galaxy
cause systematic uncertainties in the three-dimensional profile of $L_{\rm B}$ and $L_{\rm K}$.
At $15'$ ($\sim$0.1~$r_{180}$) offset from the cluster center, 
a luminous elliptical galaxy NGC~4709 exists in the middle of the 
CEN 45 field of view.
$L_{\rm K}$ of this galaxy is 40\% of that of the cD galaxy 
and causes a sudden  increase in the integrated $L_{\rm B}$ and $L_{\rm K}$ profiles.
However, the heliocentric  radial velocity of NGC 4709 is more than 1500 km/s higher than
that of the cD galaxy NGC~4696.
Distance measurements with the surface brightness fluctuation and
a globular cluster luminosity function  indicate that
NGC~4709  is located in front of the Centaurus cluster 
\citep{Mieske2005}.
Therefore, excluding NGC~4709, we deprojected the brightness profiles of $L_{\rm B}$ and
$L_{\rm K}$ and derived the integrated profiles, which are
plotted as dashed lines in figure \ref{fig:IntegratedMandL}.
The difference between the solid and dashed lines in the figure shows the systematic uncertainty caused by the uncertainty in the position of NGC~4709.

\begin{figure*}[t]
\begin{minipage}{0.33\textwidth}
\FigureFile(55mm,55mm){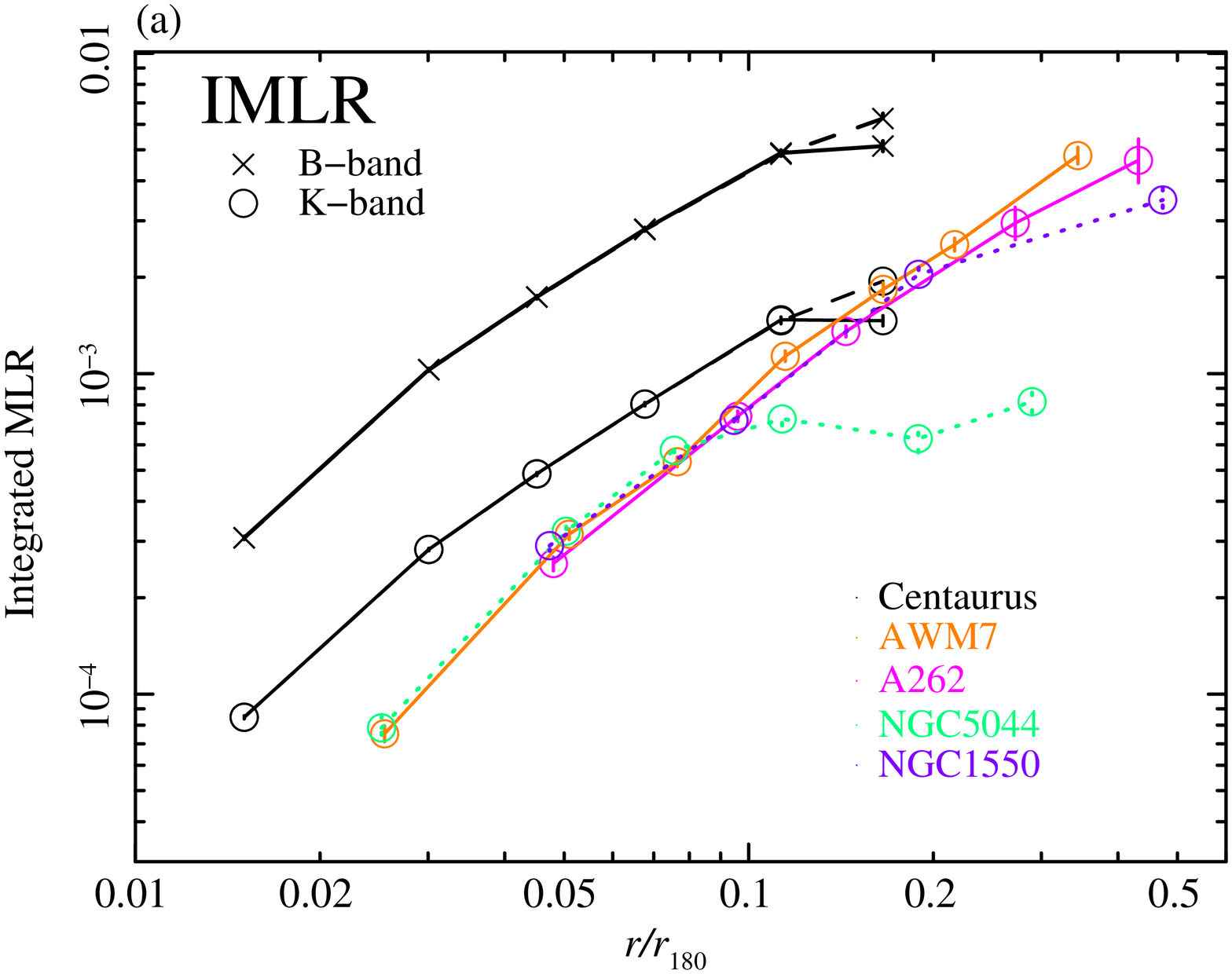}
\end{minipage}\hfill
\begin{minipage}{0.33\textwidth}
\FigureFile(55mm,55mm){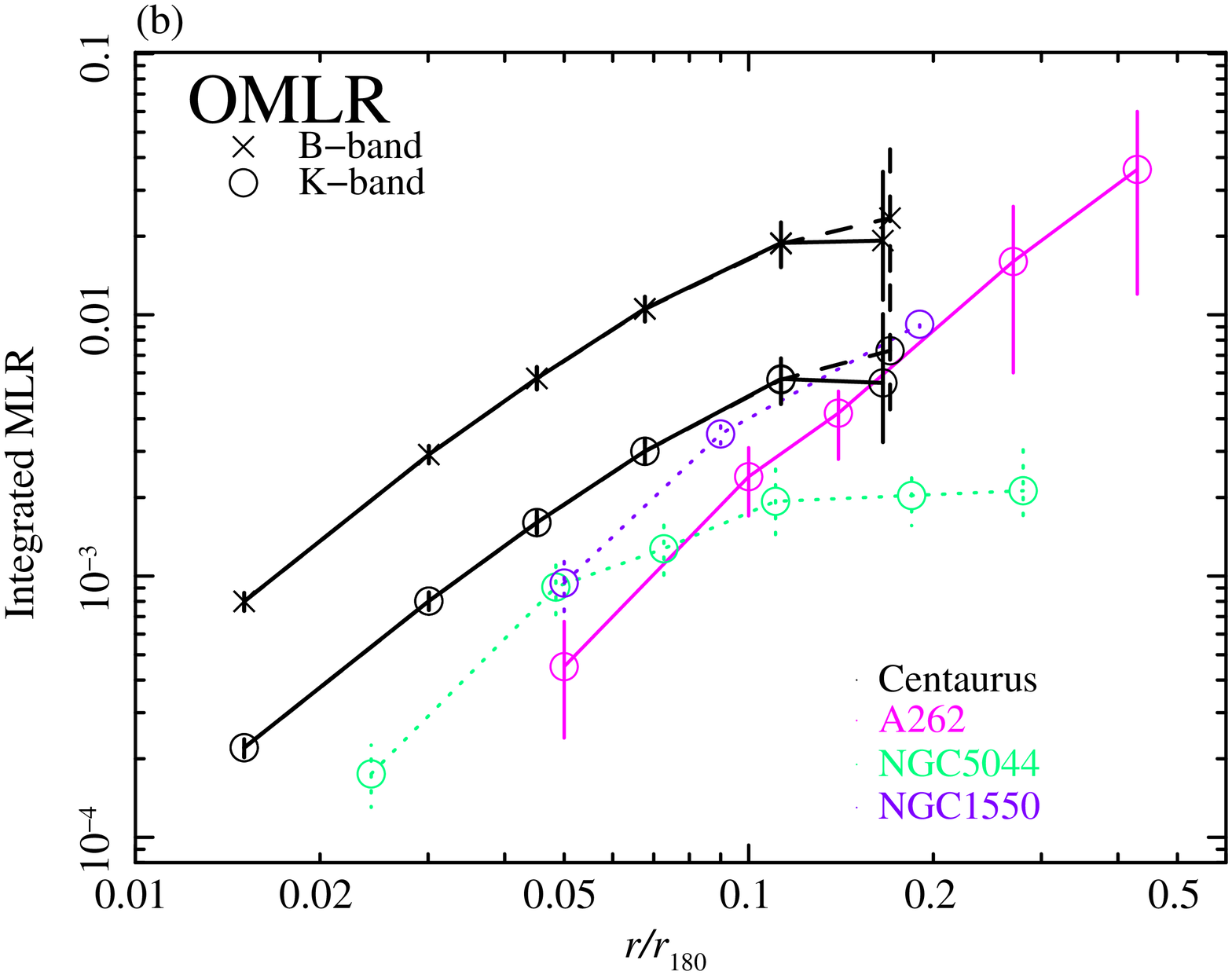}
\end{minipage}\hfill
\begin{minipage}{0.33\textwidth}
\FigureFile(55mm,55mm){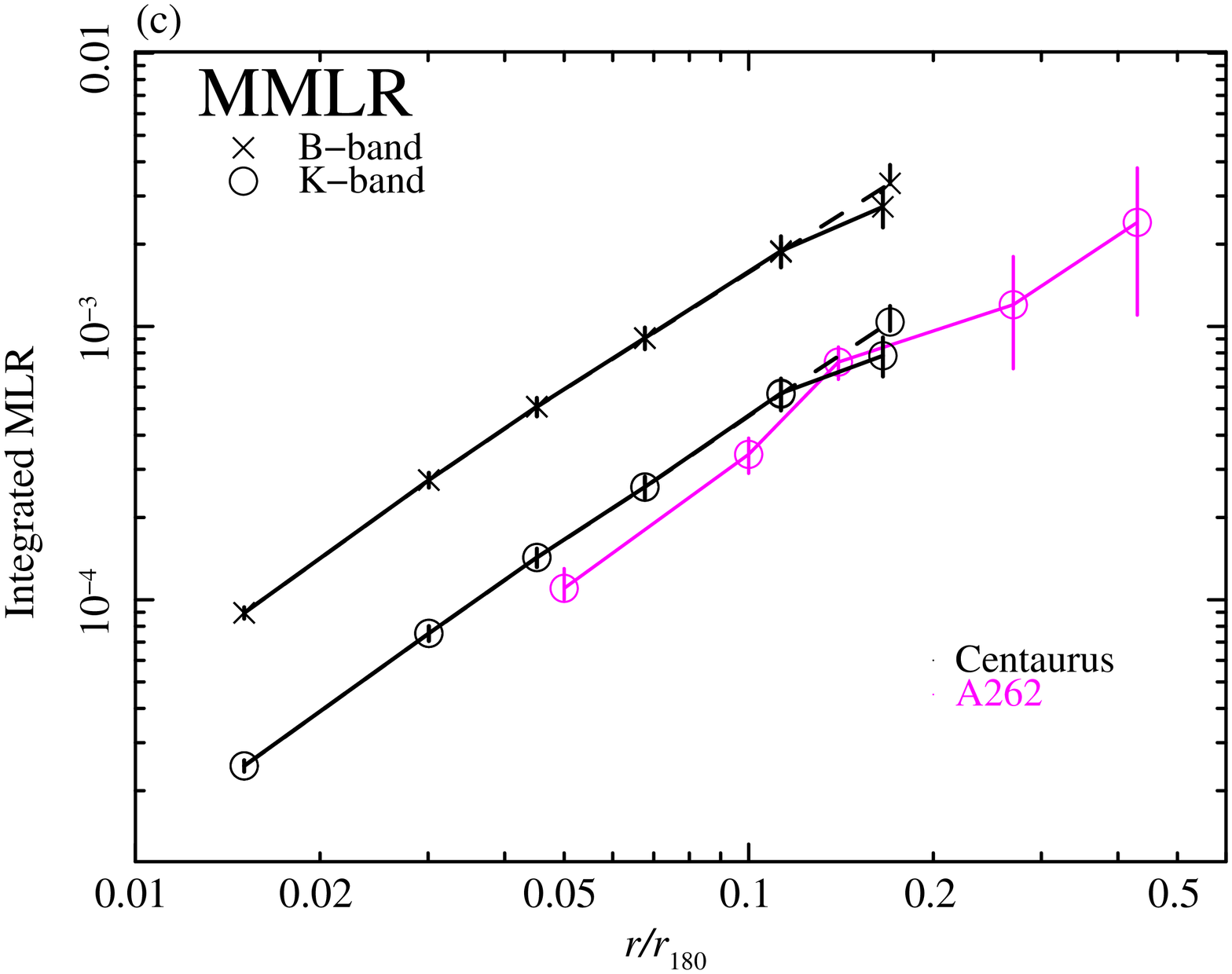}
\end{minipage}

\caption{
(a) Integrated Fe mass-to-light ratio (IMLR), (b) Oxygen
 mass-to-light ratio (OMLR), (c) Magnesium mass-to-light ratio (MMLR) 
in units of $M_{\solar} / L_{\solar}$ 
in the B-band (cross) and K-band (open circle).
Dashed line represents the result using the calculated luminosity, excluding NGC~4709.
 For comparison, we plotted the radial profile of cluster of galaxies 
AWM~7 (\cite{Sato2008}, Sato et al. in preparation) and Abell~262 \citep{Sato2009b} 
and galaxy groups NGC~5044 \citep{Komiyama2009} and NGC~1550 \citep{Sato2010}.
}
\label{fig:MLRimage}
\end{figure*}

We used the radial profile of gas mass derived from ASCA and ROSAT observations up to 22$'$ \citep{Ikebe1999} and calculated an integrated mass profile of O, Fe, and Mg 
(figure \ref{fig:IntegratedMandL}). 
We then derived 
integrated profiles of MLRs for O, Mg, and Fe (OMLR, MMLR and IMLR) 
using B-band and K-band luminosities (figure \ref{fig:MLRimage} and table \ref{tb:MLRresult}).
The error bars of the MLRs include only the abundance errors.
The IMLR, OMLR, and MMLR profiles increase with radius to $\sim 0.17~r_{180}$, 
although the profiles have a flatter slope in the 0.1--0.17 
$r_{180}$ region where NGC~4709 
is probably located in front of the Centaurus cluster.  
Excluding the galaxy, the MLR 
profiles become a smoother slope to $\sim 0.17~r_{180}$.

Figure \ref{fig:MLRimage} compares the integrated IMLR, OMLR, and MMLR profiles
of the Centaurus cluster and two other clusters Abell~262 and AWM~7
  and two groups of galaxies NGC~5044 and NGC~1550 
with dominant elliptical galaxies in their centers.
Within 0.1 $r_{180}$, where the central galaxies dominate the
integrated $L_{\rm K}$ profiles, the IMLR, OMLR, and MMLR of the Centaurus cluster tends
to be higher than those of the other systems by a factor of $1.5\sim 2$.
In contrast, at 0.1--0.17 $r_{180}$,  MLRs of the Centaurus cluster become
consistent with those of the other systems.

\begin{table*}[t]
\caption{Weighted averages of the abundance ratios in solar units.}
\label{tb:abdratiotable}
\begin{center}
\begin{tabular}{lllllll}
\hline
region (arcmin/$r_{180}$)& O/Fe & Mg/Fe & Si/Fe & S/Fe & Ar/Fe & Ca/Fe \\
\hline
0$'$--6$'$/0--0.046 & 0.58$^{+0.04}_{-0.04}$ & 0.54$^{+0.03}_{-0.03}$ & 0.79$^{+0.02}_{-0.02}$ & 0.82$^{+0.02}_{-0.02}$ & 0.94$^{+0.06}_{-0.06}$ & 1.17$^{+0.07}_{-0.07}$ \\
6$'$--15$'$/0.046--0.11 & 0.90$^{+0.18}_{-0.18}$ & 0.74$^{+0.12}_{-0.12}$ & 0.70$^{+0.07}_{-0.07}$ & 0.79$^{+0.10}_{-0.10}$ & 1.02$^{+0.22}_{-0.22}$ & 1.15$^{+0.25}_{-0.25}$ \\
15$'$--23$'$/0.11--0.17 & 0.75$^{+1.66}_{-0.75}$ & 1.43$^{+0.52}_{-0.49}$ & 0.88$^{+0.29}_{-0.28}$ & 0.83$^{+0.36}_{-0.37}$ & 1.98$^{+0.98}_{-0.96}$ & 0.97$^{+1.00}_{-0.97}$ \\
\hline
\end{tabular}
\end{center}
\end{table*}

\subsection{Metal Enrichment Histories Outside the Cool Core}

Outside the cool core, from 0.05~$r_{180}$ to $\sim$0.17~$r_{180}$,
the observed Fe abundance profile of the Centaurus cluster
agrees well with those of the weighted average of the nearby cD clusters 
observed with XMM (\cite{Matsushita2011}).
At this radial range,  the IMLR, OMLR, and MMLR of the Centaurus cluster
are also similar to those of  AWM~7,  Abell~262, and NGC~1550 observed with Suzaku,
considering the difference in the integrated values within 0.05 $r_{180}$.
The abundance patterns of O/Mg/Si/S/Fe of these systems are almost similar,
although the Mg/Fe ratio of the AWM 7 cluster at 0.05--0.1 $r_{180}$ 
tends to higher.
These results indicate that
outside the cool core regions of clusters, 
the metals in the ICM may be universal in clusters of galaxies, 
and clusters of galaxies may have universal metal enrichment histories.

The observed abundance pattern of O/Mg/Si/S/Ar/Ca/Fe beyond 0.05~$r_{180}$
is between those of 
SN II and SN Ia from nucleosynthesis models (figure \ref{fig:abdratiocomp}).
Therefore, both SN Ia and SN II products have been mixed into the ICM.
The observed solar abundance pattern from O to Fe indicates that a large 
quantity Fe 
was synthesized by SN Ia, and the number ratio of SNe II to SNe Ia is 
estimated to be 3$\sim$4 \citep{Sato2007b, Sato2008}.
The similarity of abundance pattern of these clusters and groups of galaxies
outside the cool cores 
indicates that the contributions of the two types of SN to the metals in the 
ICM are similar. 

\citet{Renzini2005} found that
the OMLR of the ICM is a very sensitive function of the slope of the initial mass
function (IMF). Adopting a Salpeter IMF, the expected value
is $\sim$0.1~$M_{\solar}/L_{\rm B,\solar}$, and
an increase in the slope of the IMF decreases the value of the OMLR.
In contrast, the integrated OMLR at 0.17 $r_{180}$ of the Centaurus cluster, 
excluding NGC~4709, is 0.01--0.04 $M_{\solar}/L_{\rm B,\solar}$.
We expect the OMLR to become 0.02--0.04 $M_{\solar}/L_{\rm B,\solar}$, assuming that 
the ratio of O and Mg is constant, since within the radius, the
abundance ratio of the O and Mg is consistent with a constant.
This value is smaller than the expected value obtained with the Salpeter IMF 
by a factor of 3--5.
Considering that  the observed MLRs of AWM~7 and Abell~262 increase with radius,
the observed values give  lower limits of the OMLRs in the entire clusters.
Therefore, to study the slope of the IMF in clusters of galaxies we need measurements of 
MLRs at outer regions of these systems.

\subsection{Metal Enrichment from the cD Galaxy}

The metals in the cool core of clusters are a mixture of those in the ICM
and the supply from the cD galaxy, which contains those synthesized by SN Ia
and those coming from stars through stellar mass loss.
The metal abundances in the ISM of elliptical galaxies provide important 
information regarding current metal supply into the ICM.
In these galaxies, the O and Mg abundances in the ISM should be equal to
those in mass-losing stars, because these elements are not substantially 
synthesized by SN Ia.
Fe abundance of the ISM is a sum of stellar metallicity and SN Ia contribution,
 which is proportional to  $M^{\rm Fe}_{\rm SN} \theta_{\rm SN}/\alpha_*$ (see
\cite{Matsushita2003} for details).  Here, $M^{\rm Fe}_{\rm SN}$ is 
Fe mass synthesized in one SN Ia, $\theta_{\rm SN}$ is the SN Ia
rate, and $\alpha_*$ is the stellar mass loss rate.

\begin{figure}[t]
\begin{center}
 \FigureFile(80mm,80mm){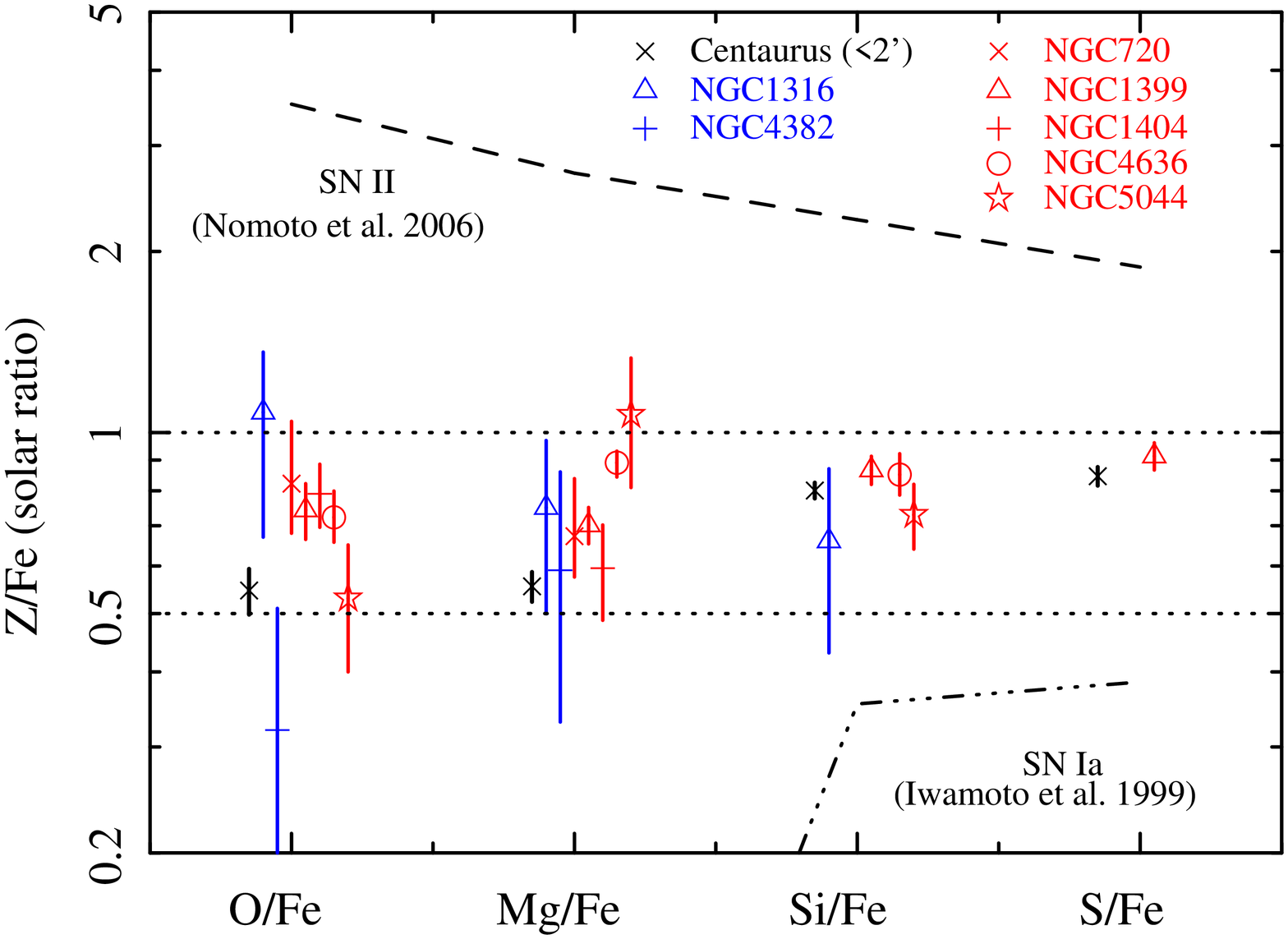}
\end{center}
\caption{
A Comparison of the abundance ratios in the $0'<r<2'$ region of the Centaurus cluster
 (black) with the ratio of S0 galaxies (blue) and elliptical galaxies (red).
S0 galaxies are two-temperature results of NGC~1316 \citep{Konami2010} 
and NGC~4382 \citep{Nagino2010}.
Elliptical galaxies are NGC~720 \citep{Tawara2008}, 
NGC~1399 \citep{Matsushita2007a}, NGC~1404 \citep{Matsushita2007a}, 
NGC~4636 \citep{Hayashi2009}, and NGC~5044 \citep{Komiyama2009}.
}
\label{fig:compES0}
\end{figure}

Figure \ref{fig:compES0} compares the central abundance pattern of the Centaurus cluster
within 2$'$  with those of early-type galaxies observed with Suzaku.
O/Fe and Mg/Fe ratios of the Centaurus cluster, 0.5--0.6 solar ratio,
are systematically smaller than those of the galaxies at 0.8--1.0 solar ratio.
The stellar metallicity of NGC~4696, the cD galaxy of the Centaurus cluster,
 calculated from the gradient of optical
 Mg$_{2}$ indexes within the effective radius is similar to those
of NGC~720, NGC~4636, and NGC~5044 plotted in figure \ref{fig:compES0} \citep{Kobayashi1999}.
Therefore, the difference in the abundance ratios may be caused by a difference 
in contribution from SN Ia, and
 the  metals in the center of the Centaurus cluster 
are not simple accumulations of hot ISM in ellipticals.
Because the abundance pattern outside the cool core regions of clusters
is similar to those of ISM in elliptical galaxies, the metals in the
center of the Centaurus cluster may not be a  simple mixture of those in the ICM and the current ISM in elliptical galaxies.

The difference between the Centaurus cluster and elliptical galaxies is the accumulation time scale of metals from stars. 
The IMLR using B-band luminosity of elliptical galaxies observed with ASCA is 
$10^{-5}$-- $10^{-4} M_{\solar}/L_{B,\solar}$ \citep{Makishima2001},
 which is a factor of 3--30 smaller than  that within 2$'$ (0.015 $r_{180}$)
of the Centaurus cluster.
The enrichment timescale of excess Fe in the center of the Centaurus cluster
is at least several Gyr \citep{Boehringer2004},
whereas the enrichment time scales of hot ISM in elliptical galaxies are smaller than 1 Gyr 
\citep{Matsushita2000, Matsushita2001}.
Therefore, a simple interpretation is 
that a longer time scale indicates that the ratio of SN Ia rate 
to stellar mass loss rate
was higher in the past.

Another interpretation is that SN Ia products in ISM in elliptical galaxies
are lost to intergalactic space by their buoyancy, as discussed in \citet{Matsushita2000}, 
based on ASCA observations of early-type galaxies.
\citet{Tang2010} simulated the evolution of hot SN Ia ejecta and
found that they quickly reach a substantially higher outward velocity than the ambient medium.
In the center of the Centaurus cluster,
these ejecta can mix with the surrounding medium because of the long enrichment time.

\begin{table*}[t]
\caption{Summary of B-band and K-band metal mass-to-light ratios of the Centaurus cluster}
\label{tb:MLRresult}
\begin{center}
\begin{tabular}{llllll}

\hline
B-band & & & & & \\
\hline
 $r$ & $L_{\rm B}$ & $M_{\rm gas}$ & IMLR & OMLR & MMLR \\
(arcmin$\slash r_{180}$) & $\times10^{11}L_{\solar}$ & 
$\times10^{11}M_{\solar}$ &
$\times 10^{-4} (M_{\solar} / L_{\solar})$ & $\times10^{-3}$ ($M_{\solar} / L_{\solar}$) 
& $\times 10^{-4} (M_{\solar} / L_{\solar})$\\
\hline
2$'\slash$0.015 & 1.5 & 0.21 & 3.07$^{+0.03}_{-0.03}$ & 0.80$^{+0.06}_{-0.06}$ & 0.89$^{+0.04}_{-0.04}$ \\
4$'\slash$0.030 & 1.5 & 0.91 & 10.3$^{+0.1}_{-0.1}$ & 3.0$^{+0.2}_{-0.2}$ & 2.7$^{+0.2}_{-0.2}$ \\
6$'\slash$0.045 & 1.5 & 2.0 & 17.4$^{+0.2}_{-0.2}$ & 5.7$^{+0.6}_{-0.5}$ & 5.1$^{+0.4}_{-0.4}$ \\
9$'\slash$0.068 & 1.5 & 4.3 & 28.2$^{+0.3}_{-0.4}$ & 10.6$^{+1.2}_{-1.1}$ & 9.1$^{+0.8}_{-0.8}$ \\
15$'\slash$0.11 & 1.8 & 11.4 & 49$^{+1}_{-1}$ & 19$^{+4}_{-4}$ & 19$^{+2}_{-2}$ \\
22$'\slash$0.17 & 2.9 & 22.0 & 51$^{+2}_{-2}$ & 19$^{+16}_{-8}$ & 27$^{+5}_{-4}$ \\
\hline
\multicolumn{2}{l}{B-band (Excluding NGC4709)} & & & & \\
\hline
15$'\slash$0.11 & 1.8 & 11.4 & 48$^{+1}_{-1}$ & 19$^{+4}_{-4}$ & 19$^{+2}_{-2}$ \\
22$'\slash$0.17 & 2.4 & 22.0 & 63$^{+2}_{-2}$ & 23$^{+20}_{-10}$ & 33$^{+6}_{-5}$ \\
\hline
\hline
K-band & & & & & \\
\hline
 $r$ & $L_{\rm K}$ & $M_{\rm gas}$ & IMLR & OMLR & MMLR \\
(arcmin$\slash r_{180}$) & $\times10^{11}L_{\solar}$ & 
$\times10^{11}M_{\solar}$ &
$\times 10^{-4} (M_{\solar} / L_{\solar})$ & $\times10^{-3}$ ($M_{\solar} / L_{\solar}$) 
& $\times 10^{-4} (M_{\solar} / L_{\solar})$\\
\hline
2$'\slash$0.015 & 5.4 & 0.21 & 0.85$^{+0.01}_{-0.01}$ & 0.22$^{+0.02}_{-0.02}$ & 0.25$^{+0.01}_{-0.01}$ \\
4$'\slash$0.030 & 5.4 & 0.91 & 2.83$^{+0.02}_{-0.03}$ & 0.80$^{+0.06}_{-0.06}$ & 0.75$^{+0.05}_{-0.05}$ \\
6$'\slash$0.045 & 5.4 & 2.0 & 4.88$^{+0.04}_{-0.06}$ & 1.6$^{+0.2}_{-0.1}$ & 1.43$^{+0.11}_{-0.11}$ \\
9$'\slash$0.068 & 5.4 & 4.3 & 8.04$^{+0.09}_{-0.10}$ & 3.0$^{+0.3}_{-0.3}$ & 2.6$^{+0.2}_{-0.2}$ \\
15$'\slash$0.11 & 6.1 & 11.4 & 14.7$^{+0.3}_{-0.3}$ & 5.7$^{+1.1}_{-1.1}$ & 5.7$^{+0.7}_{-0.7}$ \\
22$'\slash$0.17 & 10.3 & 22.0 & 14.6$^{+0.5}_{-0.5}$ & 5.5$^{+4.5}_{-2.2}$ & 7.8$^{+1.3}_{-1.2}$ \\
\hline
 \multicolumn{2}{l}{K-band (Excluding NGC4709)} & & & & \\
\hline
15$'\slash$0.11 & 6.1 & 11.4 & 14.6$^{+0.3}_{-0.3}$ & 5.6$^{+1.1}_{-1.1}$ & 5.6$^{+0.7}_{-0.7}$ \\
22$'\slash$0.17 & 7.7 & 22.0 & 19.4$^{+0.7}_{-0.7}$ & 7.3$^{+6.1}_{-3.0}$ & 10.4$^{+1.7}_{-1.7}$ \\
\hline
\end{tabular}
\end{center}
\end{table*}

The central value of Fe abundance of the Centaurus cluster
is highest among nearby cool core clusters \citep{Sanders2006, DeGrandi2009, Matsushita2011}
within 0.03~$r_{180}$, Fe abundance of the Centaurus cluster is a
factor of two higher than the average of the other clusters with cool cores.
Within the cool core, the slope of Fe abundance of the
Centaurus cluster is the steepest, while those of 
 the other  nearby clusters with the cool cores are similar \citep{Matsushita2011}.

With XMM observations of the Centaurus, Virgo, Perseus, and Abell~1795 clusters,
\citet{Boehringer2004} found that enrichment times for the central 20 kpc 
of the Centaurus cluster is a factor of two greater than those of the Virgo and
Perseus clusters and a factor of 5 greater than that of the Abell 1795 cluster.
We found that the integrated IMLR of the Centaurus cluster is a factor of two
higher than that of Abell~262 and AWM~7 clusters.
Furthermore, 
the central O/Fe and Mg/Fe ratios of the Centaurus cluster
tend to be smaller than those of the other groups and clusters of galaxies 
(figure \ref{fig:abdratiocomp}).
These results indicate that the central regions of the Centaurus cluster
contain more SN Ia yields than the other clusters.

With Chandra observations, \citet{Fabian2005} found that 
 the core of the Centaurus cluster is complicated with  bubbles and filaments,  and
sloshing motions of the gas within the central potential well or the central
active galactic nuclei may cause the central structure of the ICM.
Dynamics of the ICM in  the cool core, particularly turbulent motions, have been used
to explain the lack of cool X-ray emitting gas in the cores of galaxy clusters.
The turbulent motion in the core of the Centaurus cluster is 
loosely constrained as $<$1400 km s$^{-1}$ from the Suzaku/XIS 
analysis of Fe-line width for the Doppler broadening \citep{Ota2007} 
and $<$1100km s$^{-1}$ from the XMM/RGS observations \citep{Sanders2011}. 
\citet{Graham2006} derived the limit of 400 km s$^{-1}$ in the central 25 kpc of the cluster based on an argument that strong turbulence would smear out, via diffusion, the observed abundance gradient.
A tighter constraint will be attained with the future high-resolution 
spectrometer onboard the ASTRO-H \citep{Mitsuda2010,Takahashi2010}.
The  larger contribution from SN Ia limits the mixing in the cool core
 in the center of the Centaurus cluster, compared to that in the other cool cores.

\section{Summary and Conclusion}

 The abundance pattern of O, Mg, Si, S, Ar, Ca, and Fe are derived up to 
the outermost radius 0.17~$r_{180}$ with Suzaku observations.
 Outside the cool core ($r >$ 0.07 $r_{180}$) all elements have 
almost the same value of 0.5 solar and the abundance pattern is 
approximately 1 solar ratio. 
The integrated IMLR is similar to the other cluster of galaxies.
The similarity of abundances and MLRs indicates that clusters of galaxies
may have universal metal enrichment histories outside the cool cores.
  Within the cool core ($r <$0.05~$r_{180}$) the abundances suddenly increase
 toward the center. Si, S, Ar, Ca, and Fe abundances have a value of 1.5 $\sim$ 2 solar, while O and Mg abundances have a value of $\sim$1 solar.
The highest central Fe abundance and IMLR and smaller O/Fe and Mg/Fe ratios compared to the other clusters limit the mixing in the cool core in the Centaurus cluster.
\\

Authors thank an anonymous referee for providing valuable comments.
NO and KM acknowledge supports by the 
Ministry of Education, Culture, Sports, Science and Technology, 
Grant-in-Aid for Scientific Research Nos. 22740124, 08034192, 10009560.


\begin{thebibliography}{99}
\bibitem[Allen and Fabian(1994)]{Allen1994}
Allen, S. W. \& Fabian, A. C. 1994, \mnras, 269, 409
\bibitem[Arnaud et al.(1992)]{Arnaud1992}
Arnaud, M., Rothenflug, R. \& Boulade, O., 
Vigroux, L. \& Vangioni-Flam, E. 1992, \aap, 254, 49
\bibitem[Baldi et al.(2007)]{Baldi2007}
Baldi, M., Chiaraluce, F. \& Kl\o ve, T. 2007, \apj, 666, 835
\bibitem[B{\"o}hringer et al.(2004)]{Boehringer2004} 
B{\"o}hringer, H., Matsushita, K., Churazov, E., Finoguenov, A. 
\& Ikebe, Y.\ 2004, \aap, 416, L21 
\bibitem[De Grandi and Molendi(2002)]{DeGrandi2002}
De Grandi, S. \& Molendi, S. 2002, \apj, 567, 163
\bibitem[De Grandi and Molendi(2009)]{DeGrandi2009}
De Grandi, S. \& Molendi, S. 2009, \aap, 508, 565
\bibitem[Ezawa et al.(1997)]{Ezawa1997}
Ezawa, H., Fukazawa, Y., Makishima, K., 
Ohashi, T., Takahara, F., Xu, H. \& Yamasaki, N.Y. 1997, \apj, 490, L33
\bibitem[Evrard et al.(1996)]{Evrard1996}
Evrard, August E., Metzler, Christopher A. \& 
Navarro, Julio F. 1996, \apj, 469, 494
\bibitem[Fabian et al.(2005)]{Fabian2005}
Fabian, A. C., Sanders, J. S., Taylor, G. B. 
\& Allen, S.W. 2005, \mnras, 360, L20
\bibitem[Finoguenov et al.(2000)]{Finoguenov2000}
Finoguenov, A., David, L. P. \& Ponman, T. J. 
2000, \apj, 544, 188
\bibitem[Finoguenov et al.(2001)]{Finoguenov2001}
Finoguenov, A., Arnaud, M. \& David, L. P. 
2001, \apj, 555, 191
\bibitem[Finoguenov et al.(2002)]{Finoguenov2002}
Finoguenov, A., Matsushita, K., 
B\"ohringer, H., Ikebe, Y. \& Arnaud, M. 2002, \aap, 381, 21
\bibitem[Fukazawa et al.(1998)]{Fukazawa1998}
Fukazawa, Y., Makishima, K., Tamura, T., Ezawa, H., 
Xu, H., Ikebe, Y., Kikuchi, K. \& Ohashi, T. 1998, \pasj, 50, 187
\bibitem[Fukazawa et al.(2000)]{Fukazawa2000}
Fukazawa, Y., Makishima, K., Tamura, T., Nakazawa, K., 
Ezawa, H., Ikebe, Y., Kikuchi, K. \& Ohashi, T. 2000, \mnras, 313, 21
\bibitem[Furusho et al.(2001)]{Furusho2001}
Furusho, T. \etal\ 2001, \pasj, 53, 421
\bibitem[Graham et al.(2006)]{Graham2006}
Graham, J., Fabian, A. C., Sanders, J. S. \& Morris, R. G. 2006, \mnras, 368, 1369
\bibitem[Hayashi et al.(2009)]{Hayashi2009}
Hayashi, K., Fukazawa, Y., Tozuka, M., Nishino, S., 
Matsushita, K., Takei, Y. \& Arnaud, K. A. 2009, \pasj, 61, 1185
\bibitem[Ikebe et al.(1999)]{Ikebe1999}
Ikebe,~Y., Makishima,~K., Fukazawa,~Y., Tamura,~T., 
Haiguang,~X., Ohashi,~T. \& Matsushita,~K. 1999, \apj, 525, 58
\bibitem[Ishisaki et al.(2007)]{Ishisaki2007}
Ishisaki,~Y., \etal\ 2007, \pasj, 59, 113
\bibitem[Iwamoto et al.(1999)]{Iwamoto1999}
Iwamoto, K., Brachwitz, F., Nomoto, K., Kishimoto, N., Umeda, H.,
 Hix, W.R. \& Thilemann, F. 1999, \apjs, 125, 439
\bibitem[Jerjen and Dressler(1997)]{Jerjen1997}
Jerjen, H. \& Dressler, A. 1997, \aaps, 124, 1
\bibitem[Kalberla et al.(2005)]{Kalberla2005}
Kalberla, P. M. W., Burton, W. B., Hartmann, D., Arnal, E. M.,
 Bajaja, E., Morras, R. \& P{\"o}ppel, W. G. L. 2005, \aap, 440, 775
\bibitem[Komiyama et al.(2009)]{Komiyama2009}
Komiyama, M., Sato, K., Nagino, R., Ohashi, T. 
\& Matsushita, K. 2009, \pasj, 61, 337
\bibitem[Konami et al.(2010)]{Konami2010}
Konami, S., Matsushita, K., Nagino, R., Tashiro, M. S., 
Tamagawa, T. \& Makishima, K., 2010, \pasj, 62, 1435
\bibitem[Kobayashi \& Arimoto(1999)]{Kobayashi1999} 
Kobayashi, C. \& Arimoto, N.\ 1999, \apj, 527, 573 
\bibitem[Koyama et al.(2007)]{Koyama2007}
Koyama, K. \etal\ 2007, \pasj, 59, S23
\bibitem[Leccardi \& Molendi (2008)]{Leccardi2008}
Leccardi, A. \& Molendi, S. 2008, \aap, 487, 461
\bibitem[Lodders(2003)]{Lodders2003}
Lodders, K. 2003, \apj, 591, 1220
\bibitem[Lumb et al.(2002)]{Lumb2002}
Lumb, D. H., Warwick, R. S., Page, M. and De Luca, A. 2002, \aap, 389, 93
\bibitem[Makishima et al.(2001)]{Makishima2001} 
Makishima, K. \etal\ 2001, \pasj, 53, 401 
\bibitem[Markevitch et al.(1998)]{Markevitch1998}
Markevitch, M., Forman, W. R., Sarazin, C. L. \& 
Vikhlinin, A. 1998, \apj, 503, 77
\bibitem[Matsushita et al.(2000)]{Matsushita2000} 
Matsushita, K., Ohashi, T. \& Makishima, K.\ 2000, \pasj, 52, 685 
\bibitem[Matsushita(2001)]{Matsushita2001} 
Matsushita, K.\ 2001, \apj, 547, 693 
\bibitem[Matsushita et al.(2003)]{Matsushita2003}
Matsushita, K., Finoguenov, A. \& B{\"o}hringer, H. 2003, \aap, 401, 443
\bibitem[Matsushita et al.(2007a)]{Matsushita2007a}
Matsushita, K. \etal\, 2007a, \pasj, 59, 327
\bibitem[Matsushita et al.(2007b)]{Matsushita2007b}
Matsushita, K., B{\"o}hringer, H., Takahashi, I. \& 
Ikebe, Y., 2007b, \aap, 462, 953
\bibitem[Matsushita(2011)]{Matsushita2011}
Matsushita, K. 2011, \aap, 527, 134
\bibitem[Maughan et al.(2008)]{Maughan2008}
Maughan, B. J. \etal\ 2008, \mnras, 387, 998
\bibitem[Mieske et al.(2005)]{Mieske2005} 
Mieske, S., Hilker, M. \& Infante, L.\ 2005, \aap, 438, 103 
\bibitem[Mitsuda et al.(2007)]{Mitsuda2007}
Mitsuda, K \etal\ 2007, \pasj, 59, 1
\bibitem[Mitsuda et al.(2010)]{Mitsuda2010}
Mitsuda, K \etal\ 2010, \procspie, 7732, 29
\bibitem[Nagino et al.(2010)]{Nagino2010}
Nagino, R. \& Matsushita, K., 2010, \pasj, 62, 787
\bibitem[Nomoto et al.(2006)]{Nomoto2006}
Nomoto, K., Tominaga, N., Umeda, H., Kobayashi, C. \&
 Maeda, K. 2006, \nphysa, 777, 424
\bibitem[Ota et al.(2007)]{Ota2007}
Ota, N. \etal\ 2007, \pasj, 59, 351
\bibitem[Renzini et al.(1993)]{Renzini1993}
Renzini, A., Ciotti, L., Pellegrini, S. \& D'Ercole, A., 1993, \apj, 419, 52
\bibitem[Renzini(2005)]{Renzini2005} 
Renzini, A.\ 2005, The Initial 
Mass Function 50 Years Later, 327, 221 
\bibitem[Sanders et al.(2006)]{Sanders2006}
Sanders, J. S. \& Fabian, A. C. 2006, MNRAS, 371, 1483
\bibitem[Sanders et al.(2008)]{Sanders2008}
Sanders, J. S., Fabian, A. C., Allen, S. W., Morris, R. G.,
 Graham, J. \& Johnstone, R. M. 2008, \mnras, 385, 1186
\bibitem[Sanders et al.(2011)]{Sanders2011}
Sanders, J. S., Fabian, A. C. \& Smith, R. K. 2011, \mnras, 410, 1797
\bibitem[Sato et al.(2007a)]{Sato2007a}
Sato, K. \etal\ 2007a, \pasj, 59, 299
\bibitem[Sato et al.(2007b)]{Sato2007b}
Sato, K., Tokoi, K., Matsushita, K., Ishisaki, Y., Yamasaki, N. Y., Ishida, M. \& Ohashi, T. 2007b, \apj, 667, 41
\bibitem[Sato et al.(2008)]{Sato2008}
Sato, K., Matsushita, K., Ishisaki, Y., Yamasaki, N. Y., Ishida, M., Sasaki, S. \& Ohashi, T. 2008, \pasj, 60, 333
\bibitem[Sato et al.(2009a)]{Sato2009a}
Sato, K., Matsushita, K., Ishisaki, Y., Yamasaki, N. Y., Ishida, M. \& Ohashi, T. 2009a, \pasj, 61, 353
\bibitem[Sato et al.(2009b)]{Sato2009b}
Sato, K., Matsushita, K. \& Gastaldello, F., 2009b, \pasj, 61, 365
\bibitem[Sato et al.(2010)]{Sato2010}
Sato, K., Kawaharada, M., Nakazawa, K., Matsushita, K., Ishisaki, Y., Yamasaki, N. Y. \& Ohashi, T. 2010, \pasj, 62, 1445
\bibitem[Schlegel et al.(1998)]{Schlegel1998}
Schlegel, D. J., Finkbeiner, D. P. \& Davis, M. 1998, \apj, 500, 525
\bibitem[Smith et al.(2001)]{Smith2001}
Smith, R. K., Brickhouse, N. S., Liedahl, D. A. \& Raymond, J. C. 
2001, \apj, 556, L91
\bibitem[Takahashi et al.(2009)]{Takahashi2009}
Takahashi, I. \etal\ 2009, \apj, 701, 377
\bibitem[Takahashi et al.(2010)]{Takahashi2010}
Takahashi, T. \etal\ 2010, \procspie, 7732, 27
\bibitem[Tamura et al.(2003)]{Tamura2003}
Tamura, T., Kaastra, J. S., Makishima, K. \& Takahashi, I. 2003, \aap, 399, 497
\bibitem[Tamura et al.(2009)]{Tamura2009}
Tamura, T. \etal\ 2009, \apj, 705, L62
\bibitem[Tang and Wang(2010)]{Tang2010} 
Tang, S. \& Wang, Q.~D.\ 2010, \mnras, 408, 1011 
\bibitem[Tawara et al.(2008)]{Tawara2008}
Tawara, Y., Matsumoto, C., Tozuka, M., Fukazawa, Y., Matsushita, K. \& Anabuki, N., 2008, \pasj, 60 307
\bibitem[Taylor et al.(2006)]{Taylor2006}
Taylor, G. B., Sanders, J. S., Fabian, A. C. \& Allen, S. W. 
2006, \mnras, 365, 705
\bibitem[Tokoi et al.(2008)]{Tokoi2008}
Tokoi, K. \etal\ 2008, \pasj, 60, 317
\bibitem[Vikhlinin et al.(2005)]{Vikhlinin2005}
Vikhlinin, A., Markevitch, M., Murray, S. S., Jones, C., Forman, W. 
\& Van Speybroeck, L. 2005, \apj, 628, 655
\bibitem[Yoshino et al.(2009)]{Yoshino2009} Yoshino, T., \etal\
2009, \pasj, 61, 805 

\end{thebibliography}
\end{document}